\documentclass[preprint,twocolumn]{aastex631}
\usepackage{subfigure}
\usepackage{color}
\usepackage{xcolor}
\usepackage{amsmath}
\usepackage{mathrsfs}
\usepackage{multirow}
\usepackage{soul}
\usepackage{amssymb}
\usepackage{hyperref}
\usepackage{natbib}
\usepackage{ulem}
\usepackage{array,makecell}
\definecolor{ultramarine}{rgb}{0.8, 0.1, 0.4}

\def\be{\begin{equation}}
\def\ee{\end{equation}}
\def\bd{\begin{displaymath}}
\def\bg{\begin{gathered}}
\def\eg{\end{gathered}}
\def\ed{\end{displaymath}}
\def\ba{\begin{aligned}}
\def\ea{\end{aligned}}

\def\nms{\mathsurround=0pt}

\def\oversim#1#2{\lower 4pt\vbox{\baselineskip 0pt \lineskip 1pt
    \ialign{$\nms#1\hfil##\hfil$\crcr#2\crcr\sim\crcr}}}

\def\bh{M_{\bullet}}

\def\msun{M_{\odot}}

\def\pyr{\,yr$^{-1}$}

\def\GNC{\texttt{GNC}}
\begin{document}

\title{Co-evolution of Nuclear Star Clusters and Massive Black Holes:\\ Extreme Mass-Ratio Inspirals}
\author{Fupeng Zhang}
\correspondingauthor{FUPENG ZHANG}
\affiliation{School of Physics and Materials Science, Guangzhou
University, Guangzhou 510006, China}[0]
\email{zhangfupeng@gzhu.edu.cn}
\affiliation{Key Laboratory for Astronomical Observation and Technology of Guangzhou, 510006 Guangzhou, China}
\affiliation{Astronomy Science and Technology Research Laboratory of Department of Education of Guangdong Province, Guangzhou 510006, China}
\author{Pau Amaro Seoane}
\affiliation{Universitat Politècnica de València, Spain}
\affiliation{
Max-Planck-Institute for Extraterrestrial Physics, Garching, Germany}


\begin{abstract}
We explore extreme mass-ratio inspirals (EMRIs) in the co-evolution
of massive black holes (MBHs) and nuclear star clusters (NSCs), which host
diverse stellar populations across a wide range of masses. The dynamics
are simulated self-consistently with \GNC, which we have updated to incorporate gravitational wave orbital decay,
the loss cone of a spinning MBH, and stellar evolution. Over $12$\,Gyr, we investigate
the evolution of the NSC with a mass-growing MBH, as well as
the EMRIs of stellar black holes (SBHs), neutron stars (NSs), white dwarfs (WDs), brown dwarfs (BDs), and
low-mass main-sequence stars (MSs), along with tidal disruption events (TDEs) involving MSs, BDs, and post-MSs.
The mass growth of the MBH contributed by TDEs
is typically $\sim 10^7\msun$, $\sim 10^6\msun$, and $\sim 5\times10^4\msun$
for massive, Milky-Way-like, and smaller NSCs, respectively. Between $40\%$ and $70\%$ of
the stellar mass is lost during stellar evolution, which dominates the mass growth of the MBH
if a significant fraction of the lost mass is accreted.
The evolution of EMRI rates is generally affected by the cluster's size expansion or contraction, stellar population evolution,
MBH mass growth, and the stellar initial mass function. The EMRI rates for compact objects peak
at early epochs ($\lesssim 1$\,Gyr) and then gradually decline over cosmic time.
LISA-band ($0.1$\,mHz) EMRIs involving compact objects around Milky-Way-like MBHs tend to have
high eccentricities, while those around spinning MBHs preferentially occupy low-inclination (prograde) orbits.
In contrast, MS- and BD-EMRIs usually have eccentricity and inclination distributions that are distinct
from those of compact objects.
\end{abstract}

\section{Introduction}

An extreme-mass ratio inspiral (EMRI) occurs when a stellar object is gravitationally captured by
a central massive black hole (MBH) and continuously spirals inward through the emission of
gravitational waves (GWs)~\citep[e.g.][]{1995ApJ...445L...7H}.
Due to the strong tidal force of the MBH, such objects are expected
to be compact, e.g.,
stellar-mass black holes (SBHs), neutron stars (NSs), and white dwarfs
(WDs)~\citep[e.g.,][]{2005ApJ...629..362H,2006ApJ...645L.133H,
2006ApJ...645.1152H, 2016ApJ...820..129B, 2018LRR....21....4A, 2022MNRAS.514.3270B, 2024PhRvL.133n1401Q},
or low-mass main-sequence stars (MSs) and brown dwarfs
(BDs, with masses in the range of $0.01\msun$-$0.1\msun$)~\citep{2003ApJ...583L..21F,2019PhRvD..99l3025A}.
For MBHs with masses of $10^4\msun$-$10^7\msun$, an EMRI is a slow process that usually involves
hundreds of thousands of inspirals emitting GWs in the frequency range of
$10^{-4}$\,Hz to $0.1$\,Hz. Thus, EMRIs are one of the
main targets for future space-based GW observatories, such as the Laser Interferometer Space Antenna
(LISA)~\citep{2017arXiv170200786A,2020GReGr..52...81B, 2023LRR....26....2A},
TianQin~\citep[e.g.,][]{2016CQGra..33c5010L}, or TaiJi~\citep{2020IJMPA..3550075R}.

The extended duration of the inspiral phase is a crucial characteristic of EMRIs. Because the timescale of gravitational wave emission scales strongly with the semi-major axis ($T\propto a^4$), EMRIs spend the majority of their lifetime in the early stages of the inspiral. This leads to the concept of Early EMRIs (E-EMRIs), which can spend hundreds of thousands of years in the detection band before the final merger~\citep{AmaroSeoaneMonoOligo}. Depending on how rapidly their frequency evolves during an observation period, E-EMRIs can be classified as monochromatic (negligible frequency evolution), oligochromatic (slow frequency evolution), or polychromatic (rapid evolution near merger)~\citep{AmaroSeoaneMonoOligo}. While the merger event rate might be low, the long residence time implies that a significant steady-state population of E-EMRIs may be present in the detector band at any given time.

EMRIs form in nuclear star clusters (NSCs) around an MBH when relaxation processes (particularly two-body relaxation)
scatter their orbits into regimes where the dynamics are dominated by
GW radiation~\citep[e.g.,][]{2005ApJ...629..362H}, or when stellar objects are trapped
within an accretion disk~\citep[e.g.,][]{2005ApJ...619...30M,2011PhRvD..84b4032K,
2021PhRvD.104f3007P,2021PhRvD.103j3018P}. The GW signals from
EMRIs encode detailed information about the MBH mass, spin, and spacetime
structure~\citep[e.g.,][]{2011PhRvD..84l4060S}.
Thus, EMRIs are expected to be ideal tools for probing
stellar dynamics in NSCs, black hole physics, and theories of gravity in strong
fields~\citep[e.g.,][]{1995PhRvD..52.5707R,2013LRR....16....7G,2018LRR....21....4A,2024arXiv240108085C}.

Furthermore, the extremely large mass ratios involved when considering the inspiral of brown dwarfs (BDs) around an MBH—termed X-MRIs—lead to even longer residence times, potentially millions of years, suggesting a non-negligible population of these sources might also be detectable~\citep{2019PhRvD..99l3025A}.

Despite extensive previous studies~\citep[e.g.,][]{1995ApJ...445L...7H,1997MNRAS.284..318S, 2003ApJ...590L..29A,
2005ApJ...629..362H,2006ApJ...645L.133H,
2006ApJ...645.1152H,2010ApJ...708L..42P,2011CQGra..28i4017A,2011PhRvD..84d4024M, 2016ApJ...820..129B,
2017PhRvD..95j3012B, 2021MNRAS.501.5012R,2022MNRAS.514.3270B,
2024PhRvL.133n1401Q,2025A&A...694A.272M}, significant uncertainties remain
regarding the dynamics, evolution, and predicted properties of EMRIs.
Most of the difficulties come from the challenges of
numerically simulating the complex stellar dynamics of an NSC around an MBH.
For example, an NSC is a dense and compact cluster typically consisting of tens of millions of stellar
objects with a broad spectrum of masses. The orbital evolution is mainly driven by two-body stochastic relaxation
processes, which are tightly coupled among components of different masses~\citep[e.g.,][]{2001A&A...375..711F,
2002A&A...394..345F, 2006ApJ...649...91F, Paper1}.

For accurate simulations of EMRIs, it is also important to distinguish them from "plunge events," where objects
directly enter the loss cone region, producing only a single burst of GW emission~\citep{1995ApJ...445L...7H}.
Even near the loss cone, it is possible for gravitational encounters with background stellar populations to scatter
the objects back to less eccentric orbits, interrupting the inspiral process.

The dynamics are further complicated by the spin of the MBH~\citep{2013MNRAS.429.3155A},
resonant relaxation~\citep{RT96,2016ApJ...820..129B},
relativistic orbital precessions (leading to the so-called
"Schwarzschild barrier")~\citep[e.g.,][]{2011PhRvD..84d4024M}, and binary effects~\citep[e.g.,][]
{2018CmPhy...1...53C,2021MNRAS.501.5012R}.
Many environmental effects can contaminate EMRI signals, such as the
dynamical perturbations of a secondary MBH~\citep{2011PhRvD..83d4030Y}, background stellar
perturbations~\citep{2012ApJ...744L..20A}, disks~\citep{2025arXiv250324084Z,2011PhRvD..84b4032K},
and dark matter~\citep[e.g.,][]{2024PhRvD.110h4080D}.

As a consequence of these dynamical complexities, many previous studies
have adopted various simplifications in the dynamics of EMRIs.
For example, the dynamics of EMRIs are often estimated according to the
steady-state solution of the system~\citep[e.g.,][]
{2005ApJ...629..362H,2006ApJ...645.1152H,2016ApJ...820..129B}.
Many previous studies assume a non-evolving MBH with a fixed mass,
ignoring the evolution of stellar populations and the stellar potential across the cluster.
The differences in the estimated EMRI event rates are large; for example, SBH-EMRI event
rates range from $10^{-9}$ yr$^{-1}$~\citep{2005ApJ...629..362H}
to $10^{-6}$ yr$^{-1}$~\citep{2003ApJ...583L..21F}.

Another important source of uncertainty is that current observational constraints on the
properties of NSCs remain poor, including the mass and density distributions of stellar
objects~\citep[e.g.,][]{2020A&ARv..28....4N},
except for a few NSCs in nearby galaxies and our own Galactic Center.
Thus, many previous studies adopt an NSC structure where the influence radius of the MBH
follows the $M-\sigma$ relation~\citep{2013ARA&A..51..511K}, under the assumption of an isothermal density distribution
for galactic nuclei. However, this assumption becomes inaccurate in several scenarios, such as
when the density profile deviates from isothermal, the MBH mass is significantly smaller than the NSC
mass~\citep{2017ARA&A..55...17A, Paper2}, or the cluster exhibits rapid size expansion
over cosmic time~\citep{Paper2}.

For an accurate study of the dynamics of an NSC and the formation of EMRIs,
it is important to simulate the co-evolution of the MBH and the NSC over cosmic time, i.e., the MBH grows its mass
by accreting surrounding gaseous material, while the stellar structure of the NSC changes in response
to the mass growth of the MBH~\citep[e.g.,][]{1980ApJ...242.1232Y,2001A&A...375..711F,Paper2}.
Due to stellar evolution,
the stellar populations of an NSC should also evolve with time~\citep[e.g.,][]{2019MNRAS.484.3279P}. The
mass loss due to stellar evolution should contribute significantly to the mass growth of
the central MBH~\citep{1991ApJ...370...60M,2002A&A...394..345F}.


\defcitealias{Paper1}{Paper~I}\defcitealias{Paper2}{Paper~II}
Here we aim to improve the study of EMRIs in NSCs
based on our previously well-tested Monte Carlo method, \GNC~\citep[][Hereafter~\citetalias{Paper1} and~\citetalias{Paper2}, respectively]{Paper1,Paper2}.
\GNC~can simulate the two-body relaxation processes by solving the Fokker-Planck equations
in the space of orbital energy and angular momentum for various types of stellar objects with
a spectrum of masses. \GNC~integrates the stellar potential and adiabatic invariant theory,
allowing for the simulation of the self-consistent dynamics of an NSC with a mass-growing MBH~\citepalias{Paper2}.
For a more comprehensive study of EMRIs, in this work we additionally include recipes for GW
orbital dissipation, the loss cone of a spinning MBH, and stellar evolution processes into \GNC.
This framework enables the simultaneous modeling of EMRIs, the
evolution of NSCs, and the mass growth of MBHs within a unified physical description, a combination not achieved
in many previous studies.

As the first application of \GNC~to EMRI studies,
we adopt an idealized scenario of an isolated, gas-free NSC in a galactic nucleus.
We assume that all stars form simultaneously in
a single burst with no subsequent star formation. While our framework can, in principle, accommodate
continuous or multiple bursts of star formation, doing so introduces additional complexities and
requires specific assumptions about both the star formation history and the location of star formation
within the cluster. We do not consider resonant relaxation here, as its effects on EMRI event rates
remain controversial~\citep[e.g.,][]{2011PhRvD..84d4024M,2016ApJ...820..129B,2006ApJ...645.1152H}.
We therefore defer investigations of these more complex scenarios,
including gaseous environments, the history of star formation, and other physical processes, to future studies.



The paper is organized as follows. In Section~\ref{sec:method}, we describe how the additional recipes are included
in \GNC. In Section~\ref{sec:dyn_emris}, we perform simulations and
investigate the evolution of the NSC, the mass growth of the MBH, and the expected properties
of EMRIs in the LISA band. Discussions and conclusions
are given in Section~\ref{sec:discuss} and~\ref{sec:con}, respectively.

\section{The method}
\label{sec:method}

The Monte Carlo method is based on our recently developed and well-tested \GNC~code~\citepalias{Paper1, Paper2},
which significantly extends the pioneering frameworks established by~\citet{SM78} and~\citet{1980ApJ...239..685M}.
\GNC~can simulate the evolution of an NSC that consists of stellar objects of various types and masses.
The diffusion process of particles is simulated by solving the Fokker-Planck equations
in the space of energy and angular momentum. The core-collapse of a Plummer cluster simulated by \GNC~is consistent with
the results of previous studies. This numerical framework enables the self-consistent treatment
of the co-evolution of the NSC and the MBH.
Below, we provide a brief description of the method; more details can be found in~\citetalias{Paper1} and~\citetalias{Paper2}.

We adopt Dehnen's model~\citep{1993MNRAS.265..250D} for the initial density profile of the cluster:
\be
\rho(r)=\frac{3-\gamma}{4\pi}
\frac{M_{\rm cl}r_a}{r^\gamma(r+r_a)^{4-\gamma}}
\ee
where $M_{\rm cl}$ is the total mass of the cluster, $\gamma$ is the density profile in the inner regions of
the cluster, and $r_a$ marks the characteristic radius where the density profile changes.
Then, approximately $10^5$ Monte Carlo sample particles are generated in the space of dimensionless
energy $x=E/\sigma_0^2$ and dimensionless angular momentum $j=J/J_c(x)$ for given masses and stellar types,
where
\begin{itemize}
\item $E=\phi(r)-\frac{1}{2}v^2$ is the specific orbital energy, where $\phi(r)$ is the gravitational potential (including the stellar and MBH potential),
and $r$ and $v$ are the radial distance to the cluster's center and the velocity, respectively;
\item $J=rv_t$ is the specific orbital angular momentum, where $v_t$ is the tangential velocity;
\item $\sigma_0$ is the user-defined characteristic velocity.
We set $\sigma_0=\sqrt{m_0G/r_0}$, where $G$ is the gravitational constant.
$m_0$ is the characteristic mass, set to
$m_0=\bh$ if the MBH mass $\bh$ is fixed,
or $m_0=10^7\msun$ if not. $r_0$ is the characteristic distance given by
$r_0=3.1~{\rm pc}(m_0/4\times10^6\msun)^{-0.55}$;
\item $J_c(x)$ is the maximum angular momentum for a given orbital energy $x$.
\end{itemize}

The potential $\phi(r)$ as a function of radius $r$
is obtained self-consistently by solving the Poisson equation through iterative methods
(see Section 2.2.2 of~\citetalias{Paper2}). According to the conservation of radial action,
the particles' energies are adjusted in response to the slowly
varying stellar and MBH potential (see Section 2.3 of~\citetalias{Paper2}).

A gas reservoir formed from the mass of TDEs~\citep{Paper2} or from mass loss during stellar evolution
(see Section~\ref{subsec:stellar_evolution} and the beginning of Section~\ref{subsec:NSCs_sf}) is assumed
to be accreted by the MBH. The consumption
rate of this gas reservoir is limited by the Eddington mass accretion rate:
\be
\dot M_{\rm edd}=2.22~\msun~{\rm yr}^{-1}\times \frac{0.1}{\epsilon}\frac{\bh}{10^8\msun},
\ee
where $\epsilon=0.1$ is the radiative efficiency.
The mass growth of the MBH is then given by $\dot \bh=\dot M_{\rm edd}(1-\epsilon)$.

The simulation is performed in iterations with adaptive timesteps $\Delta t$, which is the minimum of the
two-body relaxation timescale
and the accretion timescale when modeling the mass growth of the MBH (see Equation 40 of~\citetalias{Paper2}).

The diffusion coefficients for the two-body relaxation
process—i.e., the drift ($D_E^{\rm NR}$ and $D_J^{\rm NR}$), diffusion ($D_{EE}^{\rm NR}$ and $D_{JJ}^{\rm NR}$), and cross
($D_{EJ}^{\rm NR}$) terms for energy and angular momentum, which include the effects of the stellar potential—are pre-calculated
before each iteration. Within each simulation iteration of timestep $\Delta t$,
each individual particle iteratively evolves its energy and angular momentum with smaller adaptive
timesteps $\delta t$ (Equation B4 of~\citetalias{Paper1}):
\be\ba
\delta E&=D_E \delta t +y_1\sqrt{D_{EE}\delta t},\\
\delta J&=D_J \delta t +y_2\sqrt{D_{JJ}\delta t}.
\label{eq:deltaej}
\ea\ee
where $D_E$ ($D_J$) is the sum of the drift terms including $D_E^{\rm NR}$ ($D_J^{\rm NR}$), and $D_{EE}=D_{EE}^{\rm NR}$ ($D_{JJ}=D_{JJ}^{\rm NR}$) is the diffusion term for energy (angular momentum).
$y_1$ and $y_2$ are two random numbers drawn from a standard normal distribution
with correlation $D_{EJ}^{\rm NR}/\sqrt{D_{EE}^{\rm NR}D_{JJ}^{\rm NR}}$.

A particle is removed if its energy
is so high that it moves too close to the MBH (typically $x>10^{5}-10^6$) or if its energy is too low
(typically $x\lesssim10^{-3}$) such that it can easily escape from the system. If the loss cone is included, particles with
$j<j_{\rm lc}$ are also removed, where $j_{\rm lc}$ is the
size of the dimensionless angular momentum of the loss cone. After all particles are evolved,
the self-consistent potential, density, and
phase-space distribution of $x$ and $j$ are updated for the next iteration.

Our previous works~\citepalias{Paper1, Paper2} adopted simplified assumptions, such as
fixed stellar masses (e.g., $1\msun$ for stars and $10\msun$ for SBHs), a non-spinning MBH,
and ignoring stellar evolution and GW orbital decay.
For the purpose of this study, we additionally include GW dissipation on orbits,
the modification of the loss cone size by a spinning MBH,
and stellar evolution for given initial mass functions (IMFs).
More details are provided in the following sections.

\subsection{Including orbital decay by gravitational wave radiation}

{Gravitational wave emission from a stellar object around an MBH is usually
considered for a Keplerian system consisting only of a particle and a point mass MBH, 
for both bound~\citep{1964PhRv..136.1224P} and 
unbound~\citep{1972PhRvD...5.1021H,1977ApJ...216..610T} orbits. In 
the case when the stellar potential is included, the Keplerian orbital energy of the particle 
$E_K=G\bh/r-v^2/2=E-\phi_\star(r)$ is no longer conserved, as now it depends 
on $\phi_\star(r)$ at each point along the orbit.
As a consequence, the Keplerian orbital elements (such as the semi-major axis $a_K$ and eccentricity $e_K$) vary along the orbit.}

{Since gravitational wave dissipation is most important near the pericenter,
the GW emission can still be accurately estimated using the instantaneous values of
$a_K=G\bh/(2E_K)$ and $e_K$ at the pericenter $r_p(E,J)$, where $r_p(E,J)$ is the pericenter of the 
orbit with energy $E$ and angular momentum $J$ (obtained by
solving Equation 5 of~\citetalias{Paper2}). As a result, we have:
\be\ba
e_K&=\left|\frac{J^2}{\bh G r_p(E,J)}-1\right|\\
a_K&=\frac{r_p(E,J)}{1-e_K}.
\ea
\ee}

{
$\phi_\star(r)$ is larger at inner regions of the cluster (see the bottom left panel of Figure 3 of~\citet{Paper2}). Thus, it is possible that near the pericenter, 
$\phi_\star(r)$ is large enough that 
the instantaneous Keplerian orbital energy $E_K<0$ (thus $a_K<0$ and $e_K>1$), i.e.,
the orbit is instantaneously unbound to the MBH. In other words, near the pericenter 
the particle encounters the MBH on a hyperbolic orbit. 
In this case, the resulting gravitational wave emission
differs slightly from that of a bound orbit.}

{To model EMRI events self-consistently, we calculate the GW radiation for bound
or unbound orbits separately. 
Let $D^{\rm GW}_E$  and $D^{\rm GW}_J$
be the drifts of energy and angular momentum due to GW radiation, respectively.
These GW orbital decays can be added to
the drift terms for orbital energy and angular 
momentum in Equation~\ref{eq:deltaej}.
The details of $D^{\rm GW}_E$ and 
$D^{\rm GW}_J$ are shown in Appendix~\ref{apx:GWnumerical}.
}

\subsection{Identifying EMRIs in the simulation}
\label{subsec:identify_emri}
In our simulations, a particle is safely
identified as an EMRI if its GW orbital dissipation timescale is much shorter
than the two-body relaxation timescale.
Specifically, this condition is satisfied when
\be\ba
&T_{\rm GW}(E,J) = \frac{E }{|D^{\rm GW}_E(E,J)|} \\
\le & C T_{\rm rlx}(E,J)=C {\rm min}\left(\frac{J^2}{D_{JJ}^{\rm NR}(E,J)},
\frac{E^2}{D_{EE}^{\rm NR}(E,J)}\right)
\label{eq:emri_cri}
\ea
\ee
where $T_{\rm GW}(E,J)$ is the GW orbital decay time and $T_{\rm rlx}(E,J)$ is the two-body relaxation
time for a given orbital energy $E$ and angular momentum $J$.
$D_E^{\rm GW}$ is the energy damping rate given by either Equation~\ref{eq:gw_emri} for bound orbits or
Equation~\ref{eq:gw_emri_hyper}
for unbound orbits. $D_{JJ}^{\rm NR}$ and $D_{EE}^{\rm NR}$ are the diffusion coefficients
from Equations 21 and 23 in~\citetalias{Paper2}, respectively.
$C$ is a threshold parameter specifying the ratio of the two-body timescale to the GW decay timescale.

We set $C=10^{-3}$, which is sufficiently small so that the obtained EMRI event rate converges.
This conservative threshold prevents the false identification of EMRI events.
Thresholds with $C\gtrsim 0.1$ are unreliable, as particles moving in the regime
$0.001T_{\rm rlx}\lesssim T_{\rm GW}\lesssim0.1T_{\rm rlx}$
 can possibly be scattered back to $T_{\rm GW}>0.1T_{\rm rlx}$ by the stochastic two-body relaxation process
(see Figure~\ref{fig:track} for
trajectories of particles in regimes where $10^{-3}T_{\rm rlx}<T_{\rm GW}<0.1T_{\rm rlx}$).

Particles that fall directly into the loss cone from regimes not satisfying Equation~\ref{eq:emri_cri}
are identified as plunge events. For more details on the simulated trajectories of EMRIs and plunge events,
see Section~\ref{subsec:formation_emri}.

\subsection{Loss cones around a spinning MBH}
\label{subsec:loss_cone}
The event rates of EMRIs depend critically on the size of the loss cone.
The loss cone is the regime in the space of orbital pericenter $r_p<r_{\rm p,lc}$ (or
angular momentum $j<j_{\rm lc}$) where a particle can be destroyed or swallowed by the MBH.
Generally, $r_{\rm p, lc}={\rm max}(r_{\rm ISO},r_{\rm td})$
is the maximum of the pericenter of the Innermost Stable Orbit (ISO), $r_{\rm ISO}$, and the tidal radius, $r_{\rm td}$.
In the space of angular momentum, it is $j_{\rm lc}={\rm max}(j_{\rm ISO},j_{\rm td})$,
where $j_{\rm ISO}$ and $j_{\rm td}$ are the angular momenta corresponding to
$r_{\rm ISO}$ and $r_{\rm td}$, respectively.
For compact objects, $r_{\rm p,lc}=r_{\rm ISO}$ or $j_{\rm lc}=j_{\rm ISO}$.
However, for gaseous stars, such as BDs and MSs, the additional limitation by the tidal radius must be included.

If the MBH is spinning, the size of the ISO, $r_{\rm ISO}$,
varies with the orbital inclination $i$ of the particle relative to the equatorial plane of the MBH.
Following a method similar to that in~\citet{2011PhRvD..84l4060S} (see also~\citet{2013MNRAS.429.3155A}),
we derive the pericenter of the ISO, $r_{\rm ISO}$,
for a particle in an inclined orbit around a spinning MBH. Generally,
$r_{\rm ISO}=\mathcal{W}r_g$, where $\mathcal{W}$
depends on the dimensionless spin parameter $a$ and inclination $i$.
For a Schwarzschild MBH ($a=0$), $\mathcal{W}=8$.
If $a=1$, $\mathcal{W}$ varies from $2.12$ (prograde orbits, $i=0^\circ$) to
$11.65$ (retrograde orbits, $i=180^\circ$).
For more details, see Appendix~\ref{apx:relativistic_ls} and~\ref{apx:mapping}.

In reality, the orbital orientation should evolve,
e.g., due to vector resonant relaxation~\citep{RT96}. However, the spatial orientation of the orbit
is likely to remain randomized during stochastic relaxation processes. Thus, for simplicity, we assign
an initially random orbital inclination to each particle, which remains fixed over time.

\subsection{Stellar evolution}
\label{subsec:stellar_evolution}
\begin{figure}
    \center
    \includegraphics[scale=0.7]{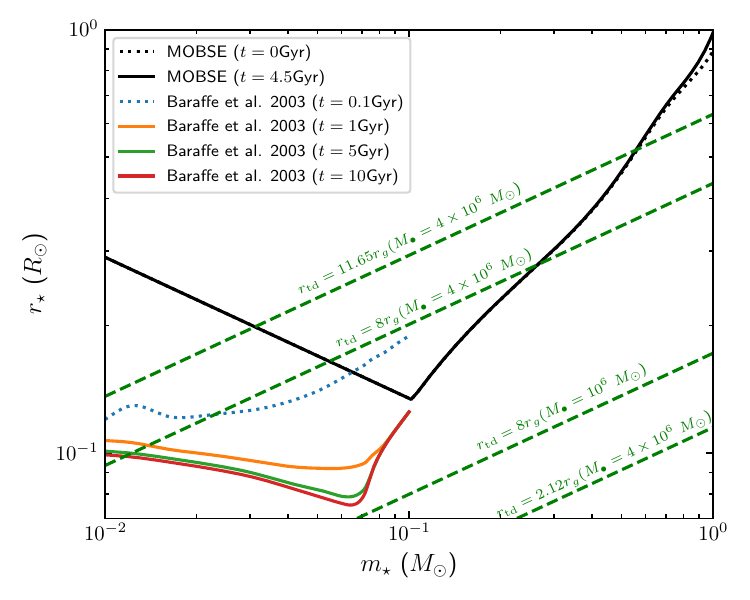}
    \caption{The mass-radius relation for BDs
    ($m_\star=0.01\msun\sim 0.1\msun$) and low-mass MSs ($0.1\msun\sim 1\msun$).
    Black lines are results from MOBSE, while
    colored lines are from~\citet{2003A&A...402..701B}. The green dashed lines show
    the stellar radius at which the tidal radius of the star, $r_{\rm td}$,
    equals the pericenter of the ISO ($r_{\rm ISO}$). For prograde (retrograde)
    orbits around a maximally spinning MBH,    $r_{\rm ISO}=2.12r_g$ ($r_{\rm ISO}=11.65r_g$).
    For a Schwarzschild MBH, $r_{\rm ISO}=8r_g$.}
    \label{fig:bd_mr}
\end{figure}

Real NSCs consist of stellar objects of multiple types, including
MSs, post-main-sequence stars (post-MSs), and compact objects.
To accurately simulate an NSC's evolution, we incorporate stellar evolution using
MOBSE~\citep{2018MNRAS.474.2959G,2018MNRAS.480.2011G}, which is an upgraded version of
BSE~\citep{2000MNRAS.315..543H,2002MNRAS.329..897H}. The stellar evolution in MOBSE is the same as that in
BSE but includes up-to-date equations for metal-dependent stellar winds,
new prescriptions for core-collapse supernovae, and a dependence of stellar winds on
the Eddington factor. Moreover, MOBSE extends the mass range of stars from $100\msun$ to $150\msun$.
For simplicity, in this study we ignore any natal kicks for NSs and SBHs.

We assume a single-starburst formation scenario, where all stellar objects begin as zero-age MSs.
The evolution of each MS is pre-computed using MOBSE over a $12$\,Gyr period.
During the simulation, the stellar type and radius of each particle are updated
according to its evolutionary table.

The stellar evolution results in a substantial population of post-MSs,
including objects in the phases of the Hertzsprung gap, first giant branch,
core helium burning, and first and second asymptotic giant branches.
For simplicity, we collectively classify them as "Post-MSs" and do not distinguish individual subtypes
in our simulation.
The evolution also generates various naked helium stars.
However, as their total number is negligible, we include them in the simulation
without further analysis. Finally, although MOBSE resolves multiple types of WDs
(Helium, Carbon/Oxygen, and Oxygen/Neon WDs), we group all of them
together as "WDs," as we are not currently interested in
investigating individual WD subtypes.

To accurately track the evolution, the simulation timestep must resolve key evolutionary
transitions. Specifically,
if $\delta t$ is the update timestep for each particle (see Appendix B in~\citetalias{Paper1}), then
 we impose an additional constraint $\delta t\rightarrow \min(\delta t, t_{\rm nk}-t)$, where
 $t$ is the current time of the particle and
 $t_{\rm nk}$ is the next key evolutionary transition time from the pre-calculated table.

\subsection{The mass-radius relation and IMF of brown dwarfs}
We note that both MOBSE and BSE use an oversimplified mass-radius relation for BDs, i.e.,
$r_\star\propto m_\star^{-1/3}$ for
$m_\star<0.1\msun$~\citep{2000MNRAS.315..543H}. The stellar radii of BDs are overestimated
by a factor of about $2-3$ compared to those in~\citet{2003A&A...402..701B} (see Figure~\ref{fig:bd_mr}).
As the rate of EMRIs is sensitive to the stellar radius of BDs, such an overestimation
of the stellar radius increases the probability of their tidal disruption and thus suppresses
the formation rate of BD-EMRIs. To improve the modeling of BD-EMRIs,
we update each BD's radius in the simulation according to
the age-dependent radius-mass relation for BDs from~\citet{2003A&A...402..701B}.

For simplicity, in this work we assume that the IMFs of BDs and MSs are continuous
around $\sim 0.1\msun$. In this case, the number ratio of BDs to
MSs for all models in this work is $\sim 0.9$. We note, however, that
BDs may constitute a distinct population from MSs~\citep{2013pss5.book..115K}, resulting in
a discontinuous IMF between the two. If we instead adopt such an IMF for
BDs, the number ratio of BDs to MSs decreases to $\sim 0.2$.
Thus, our simulated event rates for BD-related phenomena, such as EMRIs
or tidal disruption events (TDEs) of BDs, would be reduced by a factor of $\sim 4.5$.
Nevertheless, since BDs contribute only a small fraction of the total cluster mass ($\sim 5\%$),
the majority of the simulation results, including the evolution of NSCs, the mass growth of the MBH, and
EMRIs of other objects, are not significantly affected by adopting a continuous or discontinuous IMF for BDs.

\section{The dynamics of EMRIs in evolving NSCs}
\label{sec:dyn_emris}
\begin{table*}
\def\Hi{-12pt}
\def\He{24pt}
\caption{Models with discrete mass components}
    \centering
\begin{tabular}{|c|c|c|c|c|c|c|c|c|c|c|}\hline
    Name &  $r_a^a$ &$\gamma^a$ &$r_{\rm eff,i}^b$ & $r_{\rm eff,f}^c$ & $r_{\rm h,f}^c$ &
    $\bh^{d}$ ($\msun$) & Components$^{e}$   \\
\hline
\rule[\Hi]{0pt}{\He} M2
& $2.17$ & $1$ & $3.9$ & $6.7$ & $3.9$
&  \multirow{5}{*}{$4\times 10^6$, Fix}
    &
    \begin{tabular}{c} MSs~($1\msun$)
    \\SBHs~($10\,\msun$),$f_\bullet=0.001$
    \end{tabular}       \\
\cline{1-6} \cline{8-8}
\rule[\Hi]{0pt}{\He} M5        & $2.17$ & $1$ & $3.9$ & $6.7$ & $3.8$
&
    & \multirow{5}{*}{
    \begin{tabular}{c} MSs~($1\msun$)
    \\SBHs~($10\,\msun$),$f_\bullet=0.001$ \\ BDs~($0.05\msun$),$f_{\rm BD}=0.9$ \\
    WDs~($0.6\,\msun$),$f_{\rm WD}=0.1$ \\ NS~($1.4\,\msun$),$f_{\rm NS}=0.01$
    \end{tabular}}       \\
\cline{1-6}
\rule[\Hi]{0pt}{\He} M5\_2      & $2.91$ & $1.5$ & $3.9$ & $6.6$ & $3.8$ &
    &  \\
\cline{1-7}
\rule[\Hi]{0pt}{\He} M5G82      & $1.5$  & $1$  & $2.7$   & $4.4$& $2.7$  & $10^4$, Growth  &         \\
\hline
\end{tabular}
%
    \tablecomments{
$^{a}$. $r_a$ (in units of pc) and $\gamma$ are the characteristic radius and the inner density
profile of Dehnen's model, respectively;\\
$^{b}$. $r_{\rm eff,i}$ (in units of pc) is the initial effective radius of the cluster;\\
$^{c}$. $r_{\rm eff,f}$ and $r_{\rm h,f}$ (both in units of pc) are the effective radius and
the influence radius of the MBH at $12$ Gyr, respectively;\\
$^d$. The initial mass of the MBH. "Fix" indicates that the MBH mass is fixed to the initial value;
"Growth" indicates that the MBH can grow due to: tidal disruption of stars; direct
swallowing of stellar objects falling into the loss cone, including EMRIs and plunge events. \\
$^e$. $f_\bullet$, $f_{\rm BD}$, $f_{\rm WD}$ and $f_{\rm NS}$ are
the number ratios of SBHs, BDs, WDs, and NSs to MSs, respectively; \\
}    %
\label{tab:modelnoSE}
%
\end{table*}

\begin{figure*}
    \center
    \includegraphics[scale=0.7]{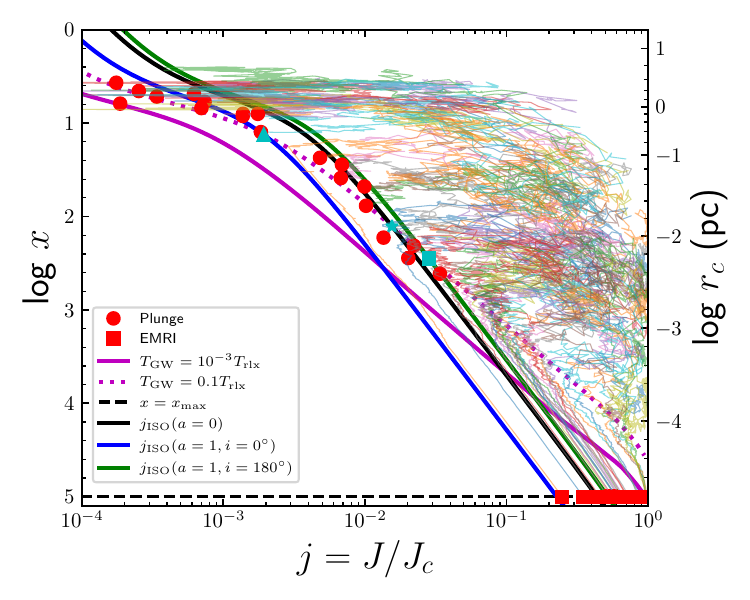}
    \includegraphics[scale=0.7]{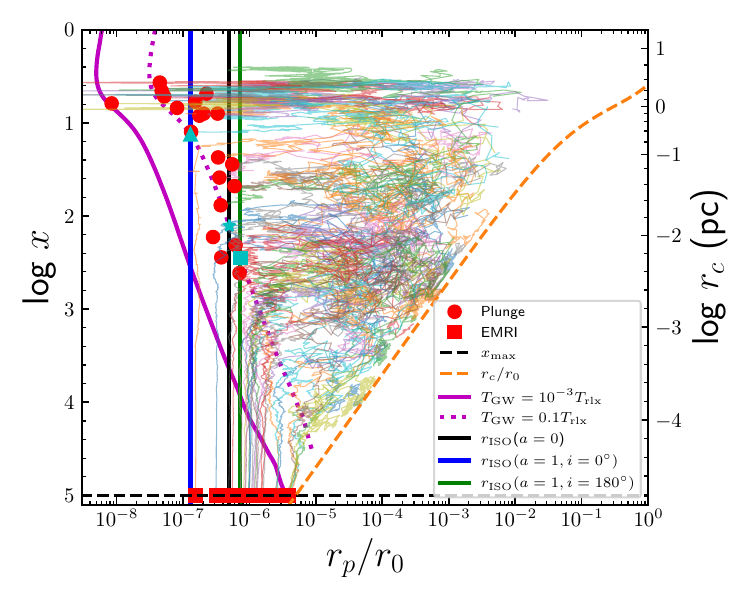}
    \includegraphics[scale=0.7]{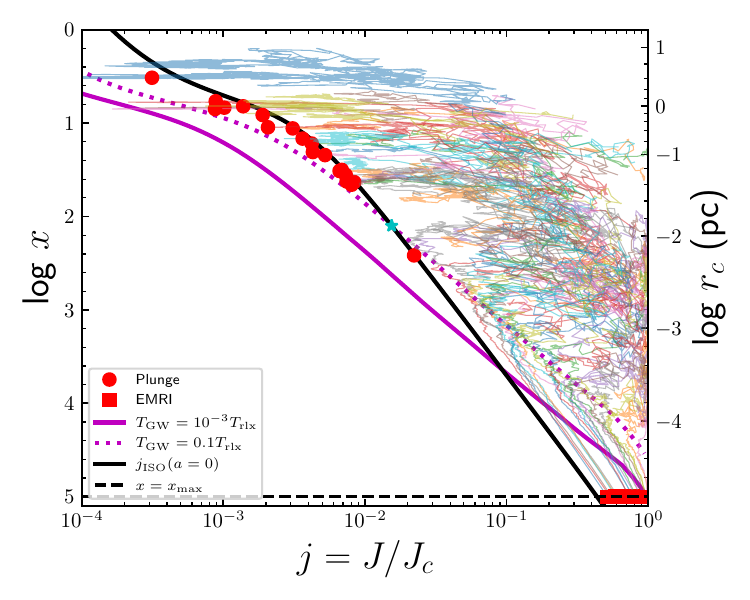}
    \includegraphics[scale=0.7]{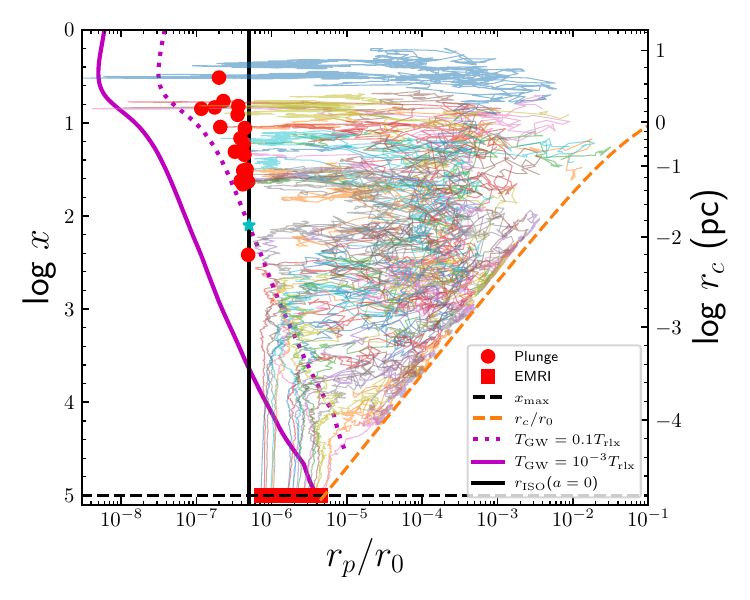}
    \caption{Examples of evolutionary trajectories of SBHs in model M2 (see Table~\ref{tab:modelnoSE}) simulated by \GNC, plotted
    in the space of energy, angular momentum, and pericenter distance.
    Thin lines ending with filled red circles
    mark plunge events, while those ending with filled red squares mark EMRI events.
    Top left panel: The energy and angular momentum evolution for a maximally spinning MBH.
    $x=E/\sigma_0^2$ and $j=J/J_c(x)$ are the dimensionless energy and angular momentum (see text in
    Section~\ref{sec:method} for their definitions). The loss cone size
    depends on the amplitude ($a$) and inclination angle ($i$) of the MBH's spin
    (defined in Equation~\ref{eq:spin_inc}).
    Top right panel: Similar to the left panel but for the evolution in the space of energy and pericenter distance. $r_0=3.1$\,pc is
    a characteristic distance. Bottom panels are similar to the top panels but for the case of a
    Schwarzschild MBH ($a=0$). The cyan filled triangle, star, and square in each panel mark the
    intersection between $T_{\rm GW}=0.1T_{\rm rlx}$ and the loss cone for $a=1$, $a=0$, and
    $a=-1$ (equivalent to $a=0$ with $i=180^\circ$), respectively.}
    \label{fig:track}
\end{figure*}

\begin{figure*}
\center
\includegraphics[scale=0.6]{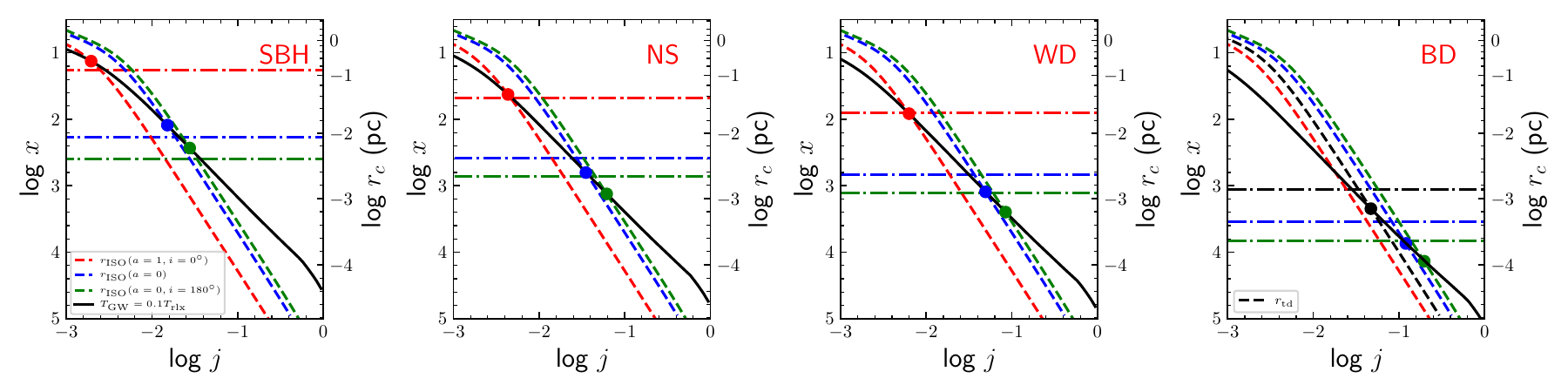}
\caption{Critical boundaries in the space of dimensionless energy $x$ and angular momentum $j$ for different
stellar objects in Model M5. In all panels, black solid lines correspond to $T_{\rm GW} =0.1T_{\rm rlx}$,
where $T_{\rm rlx}$ and $T_{\rm GW}$ are the relaxation and GW timescales, respectively
(defined in Equation~\ref{eq:emri_cri}). Red, blue, and green dashed lines indicate the pericenter of the ISO:
$r_{p} =r_{\rm ISO}=2.12r_g$ (for $a=1$ and $i=0^\circ$), $r_{\rm ISO}=8r_g$ (for $a=0$), and $r_{\rm ISO}=11.6r_g$
(for $a=1$ and $i=180^\circ$), respectively. The black dashed line
in the right panel represents the tidal radius of a BD ($r_{p}=r_{\rm td}=4r_g$)
for an MBH mass of $4\times10^6\msun$.
The loss cone size is given by $r_{p,\rm lc}={\rm max}(r_{\rm ISO},r_{\rm td})$, and its intersection with
 $T_{\rm GW}=0.1T_{\rm rlx}$ defines the critical distance $a_{\rm crit}$
(marked by red, blue, green, or black filled circles in each panel).
Horizontal dash-dotted lines, colored the same as the filled circles,
show the corresponding $a_{\rm crit}$ given by Equation~\ref{eq:emri_cir_ana}.
}
\label{fig:agw}
\end{figure*}

\begin{figure*}
    \center
    \includegraphics[scale=0.75]{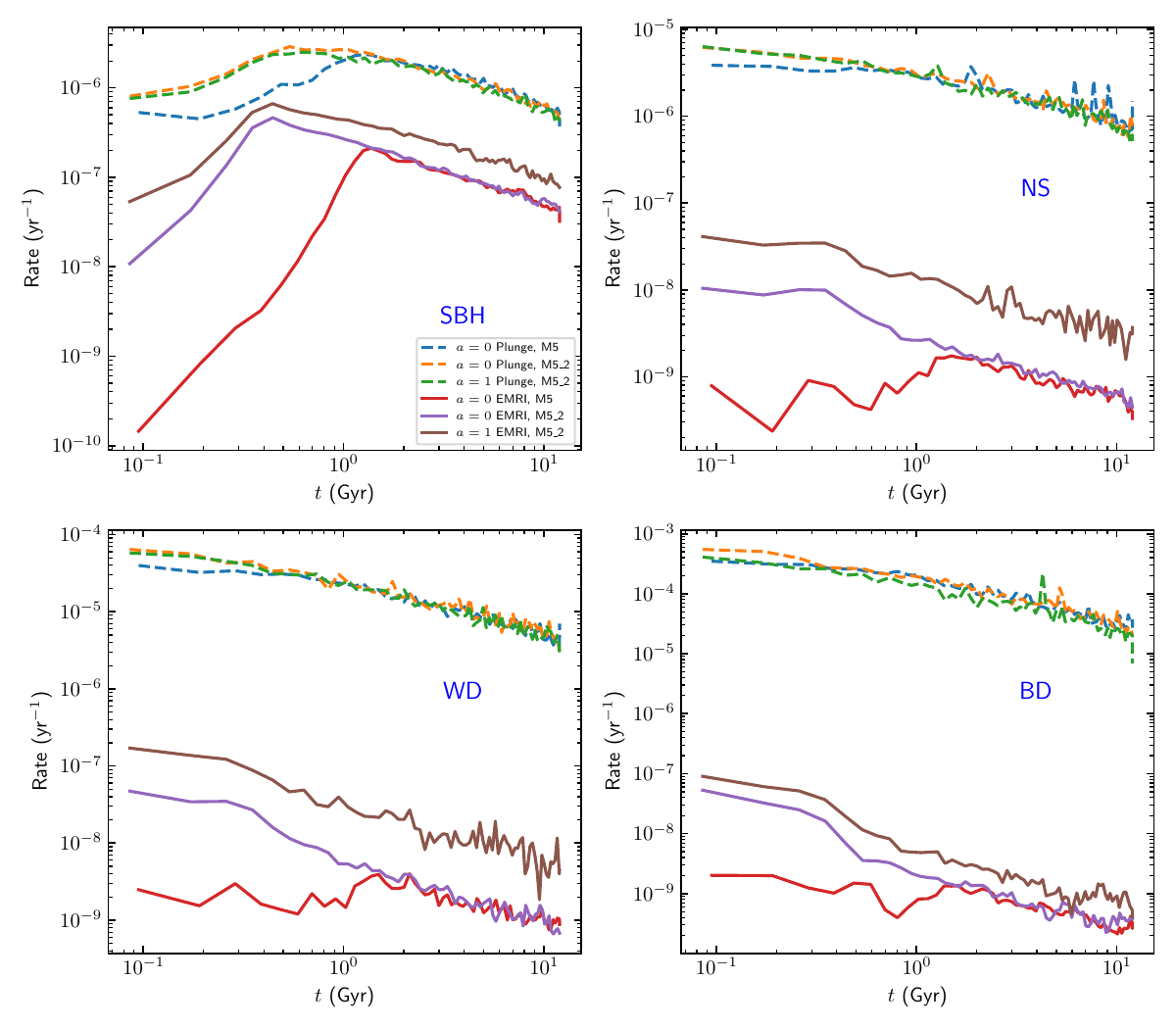}
\caption{Evolution of the EMRI and plunge event rates in five-component models
(model M5 or M5\_2 in Table~\ref{tab:modelnoSE}), with a Schwarzschild ($a=0$) or maximally spinning MBH ($a=1$).
In these simulations,
the MBH mass is fixed, and stellar evolution is ignored.
}
\label{fig:emris_lc5}
\end{figure*}

Simulations incorporating all the aforementioned recipes can provide a more comprehensive study of the evolution of
NSCs, the mass growth of the MBH, and the dynamics of EMRIs.
However, the coupling between these recipes substantially complicates the analysis of the simulation results.
For a clear analysis of EMRI dynamics, we first adopt simplified models in Section~\ref{subsec:fix_component}, assuming that
the NSC consists of either two or five mass components while neglecting stellar evolution.
These simplified scenarios facilitate the analysis of the dynamics in later sections and
comparison with many previous studies that adopted similar assumptions.
We extend our analysis later in Section~\ref{subsec:NSCs_sf} to more realistic scenarios, where the NSC consists
of a spectrum of mass components and includes stellar evolution.

\subsection{NSCs with fixed mass components and no stellar evolution}
\label{subsec:fix_component}
In this section, we ignore stellar evolution and focus on NSC models composed of either two or five components.
The initial conditions of the models are shown in Table~\ref{tab:modelnoSE}. All models adopt
$M_{\rm cl}=4\times10^7\msun$, similar to that of our Milky Way's NSC.
In the two-component model (M2), the NSC consists of $1\msun$ MSs and
$10\msun$ SBHs, where SBHs have a number fraction of $0.001$ relative to MSs.
In the five-component models (M5, M5\_2, or M5G82), we additionally include $0.05\msun$ BDs with a number
fraction of $0.9$ relative to MSs, $0.6\msun$ WDs with a fraction of $0.2$, and $1.4\msun$ NSs with a fraction of $0.01$.
The fraction of BDs follows the results of $12$\,Gyr of continuous star formation with a Kroupa IMF~\citep{2001MNRAS.322..231K},
as shown in Figure 9 and Table 2 of~\citetalias{Paper1}. The radius of the BDs is $0.08R_\odot$ according to
\citet{2003A&A...402..701B} for a $0.05\msun$ BD with an age of $>5$\,Gyr.

We first focus on models M2, M5, and M5\_2 (Sections~\ref{subsec:formation_emri},~\ref{subsec:critic_a}, and~\ref{subsec:fix_mass_emri_rate}), all of which adopt a fixed MBH mass of
$4\times10^6\msun$, similar to the MBH in our Galactic Center~\citep{2009ApJ...692.1075G}.
We then investigate the slightly more complicated model M5G82
(Section~\ref{subsec:fix_comp_mass_growth}), in which the MBH mass grows from $10^4\msun$
by swallowing objects that fall
into the loss cone (including EMRIs and plunge events) and by accreting gas released by MS-TDEs.
Details of the accretion processes can be found in Section 4.2 of~\citetalias{Paper2}.
For all models, the evolution and rates of TDEs are similar to those in~\citetalias{Paper2}, and thus we will not
discuss them further here.

\subsubsection{The formation of EMRIs}
\label{subsec:formation_emri}
Figure~\ref{fig:track} shows the
evolution of SBHs in the space of dimensionless energy ($x$), angular momentum ($j$), and
 pericenter distance $r_p/r_0$ ($r_0=3.1$\,pc) for the two-component model (M2)
around a Schwarzschild or a maximally spinning MBH.
It is apparent that below the boundary defined by
Equation~\ref{eq:emri_cri}, the evolution is dominated by GW radiation, and the SBHs eventually
become EMRIs.

The reduced pericenter distance of ISOs around a maximally spinning MBH
—reaching as low as $2.12r_g$ compared to the $8r_g$ of
a Schwarzschild black hole—significantly expands the spatial regime
within the cluster where particles can become EMRIs.
This enhanced parameter space particularly affects particles in the cluster's outer regions,
enabling them to transition into EMRIs despite their larger orbital semi-major axes.
As a consequence, we expect that the rates of EMRIs will be higher around a spinning MBH.

SBHs entering the loss cone from regimes above the boundary
become plunge events. It is apparent from the right panels of Figure~\ref{fig:track}
that they fall into the MBH with a pericenter distance smaller than $r_{\rm ISO}$ (the pericenter of the ISO).
Note that a small fraction of plunge events exhibit multiple decreases in orbital energy
due to GW emission before falling into the loss cone.
However, they are not identified as EMRIs because their dynamics are still dominated by the relaxation
process rather than by GWs, i.e., they do not satisfy Equation~\ref{eq:emri_cri}.

\subsubsection{The critical radius $a_{\rm crit}$ for different stellar objects}
\label{subsec:critic_a}
The intersection between $T_{\rm GW}\sim 0.1T_{\rm rlx}$\footnote{Note that here we use $C\sim 0.1$
so that the majority of EMRI events are identified within a relaxation timescale.
This is different from the value $C=0.001$
described in Section~\ref{subsec:identify_emri}, which is the critical value to
identify a particle as an EMRI for an instantaneous momentum in the simulation.}
and the loss cone size, $j=j_{\rm lc}$,
defines a critical radius $a_{\rm crit}$ (or energy $E_{\rm crit}$) below (or above)
which particles are likely to become EMRIs within their
relaxation time~\citep[e.g.,][]{2006ApJ...645L.133H}.
The critical radius $a_{\rm crit}$ corresponds to
the radius of a circular orbit $r_c(E_{\rm crit})$ for the critical
energy $E_{\rm crit}$. If the stellar potential is ignored,
$a_{\rm crit}=G\bh/(2E_{\rm crit})$, which reduces to the Keplerian orbital semi-major axis.
The larger the value of $a_{\rm crit}$, the higher the expected rates of EMRIs.

The critical radius $a_{\rm crit}$ for different stellar objects
can be obtained from our simulation.
In Model M5, the positions of $a_{\rm crit}$ for various types of stellar objects are shown
in Figure~\ref{fig:agw}.
For prograde orbits around a maximally spinning MBH,
$a_{\rm crit}\sim 210$\,mpc, $43$\,mpc, $20$\,mpc, and $0.7$\,mpc
for SBHs, NSs, WDs, and BDs, respectively.
These values are $\sim 10$ times larger than those for a Schwarzschild MBH
and $\sim 20$ times larger than in the case of retrograde orbits around a maximally spinning MBH.
Thus, assuming initially randomized stellar orbit orientations,
the EMRI rates around a spinning MBH are expected to be higher than around a Schwarzschild MBH.

Note that for BDs, the loss cone size is limited by their tidal radius.
For example, in the case of $a=1$ and $i=0^\circ$, we still have
$r_{\rm p,lc}=r_{\rm td}=4r_g>r_{\rm ISO}=2.1r_g$,
as shown in the right panel of Figure~\ref{fig:agw}.
As a consequence,
the enhancement of BD-EMRI event rates around a maximally spinning MBH is not as
significant as for other compact objects.

Note that the smaller the MBH mass, the smaller the value of $a_{\rm crit}$ for BDs.
When $\bh\lesssim 10^5\msun$, BDs become EMRIs only if they can
be inside the ISO. In these cases, none of the BDs can become EMRIs.

We can compare $a_{\rm crit}$ from our simulation with results from theoretical analysis, which
is approximately given by~\citep{2019PhRvD..99l3025A}:
\be\ba
a_{\rm crit}&=\epsilon R_0
\left(\mathscr{W}^{5/2}N_\bullet\ln\Lambda
\frac{m_{\rm SBH}^2}{\bh m_{\rm EMRI}}
\right)^{\frac{1}{\beta-3}}\\
\epsilon&=\left[16.97C(3-\beta)(1+\beta)^{3/2}
\right]^{1/(\beta-3)}
\label{eq:emri_cir_ana}
\ea\ee
where
\begin{itemize}
\item $m_{\rm EMRI}$ is the mass of the EMRI object;
\item $m_{\rm SBH}$ is the mass of the SBHs;
\item $\mathscr{W}=r_{p,\rm lc}/(8r_g)={\rm max}(r_{\rm ISO}, r_{\rm td})/(8r_g)$.
If $r_{p,\rm lc}=r_{\rm ISO}$, it depends on the amplitude $a$ and inclination $i$ of the MBH's spin;
\item $R_0$ is the radius within which dynamics are dominated
by SBHs, obtained by solving $\rho_\star(R_0)m_\star=\rho_\bullet(R_0)m_{\rm SBH}$,
where $\rho_\star$ and $\rho_\bullet$ are the mass density distributions of stars and SBHs;
\item $N_\bullet$ is the number of SBHs within $R_0$;
\item $C=0.1$; $\Lambda=\bh/\msun$.
\item $\beta$ is the density slope of the objects contributing to EMRIs.
According to~\citetalias{Paper1}, we can approximately adopt $\beta=1.75$ for SBHs and $\beta=1.5$ for
other compact objects.
\end{itemize}

The precise values of both $R_0$ and $N_\bullet$ in Equation~\ref{eq:emri_cir_ana} can be obtained from
our numerical simulations. For model M5, we find that at $12$\,Gyr,
$R_0=0.26$\,pc and $N_\bullet=1946$. Adopting these values, we can estimate
$a_{\rm crit}$ for different stellar objects from
Equation~\ref{eq:emri_cir_ana}. The values for SBHs, NSs, WDs, and BDs are shown as horizontal dash-dotted lines in the
panels of Figure~\ref{fig:agw}. We can see that the simulation results are roughly consistent
(within about one order of magnitude) with the analytical estimations.

\subsubsection{The evolution of the event rates}
\label{subsec:fix_mass_emri_rate}
Figure~\ref{fig:emris_lc5} shows the evolution of EMRI and plunge event rates in models M5 and M5\_2,
for both a spinning and a non-spinning MBH.
The EMRI rates depend on the spin of the MBH and exhibit significant
evolution over cosmic time.

For SBHs, the EMRI rates rise and peak around $0.4\sim 1$\,Gyr with values of $2.18\sim6.88\times10^{-7}$\pyr,
then gradually decline to $\sim (4.7-8.3)\times10^{-8}$\pyr~by $12$\,Gyr,
mainly due to the depletion of SBHs and the size expansion of the cluster.
The plunge event rates for SBHs follow a similar evolution but with much higher peak rates
of $\sim2\times10^{-6}$\pyr.
Such a rise-and-fall behavior of SBH-EMRIs and plunge events—when the mass growth of the MBH is neglected—is
consistent with the results in~\citet{2022MNRAS.514.3270B}.

For NSs, WDs, and BDs, their initial rates before $0.1$\,Gyr are $(0.1-4)\times10^{-8}$\pyr, $(0.1-20)\times10^{-8}$\pyr,
and $(0.1-10)\times10^{-8}$\pyr, respectively, but then decrease gradually to $(0.8-3)\times10^{-9}$\pyr,
$(0.9-4)\times10^{-9}$\pyr, and $(2-8)\times10^{-10}$\pyr~by $12$\,Gyr, respectively.

These results can be understood from the discussion of $a_{\rm crit}$
in Section~\ref{subsec:critic_a}. As
the $a_{\rm crit}$ of NSs, WDs, and BDs are much smaller than that of SBHs, their
rates of EMRIs are orders of magnitude lower.
For all compact objects, the EMRI rates are generally
$\sim 3-4$ times higher around a maximally spinning MBH than around a non-spinning MBH.

Theoretically, the rates can be estimated using Equation 32 of~\citet{2019PhRvD..99l3025A}:
\be
\ba
\dot\Gamma_{\rm EMRI}&\simeq\frac{3-\beta}{2\lambda}\frac{N_{\rm obj,0}}{T_0R_0^\lambda}
a_{\rm crit}^\lambda\left[\ln(\Lambda_{\rm crit})-\frac{1}{\lambda}\right]
\label{eq:emri_rate_ana}
\ea\ee
where
\begin{itemize}
\item $\lambda=9/2-\beta-\gamma$, where $\gamma$ is the power-law index of the number density of SBHs;
\item $\Lambda_{\rm crit}=a_{\rm crit}/(8R_s)$, where $R_s$ is the Schwarzschild radius;
\item $N_{\rm obj,0}$ is the number of EMRI objects within $R_0$;
\item $T_0$ is the two-body relaxation time at $R_0$.
\end{itemize}
The definitions of $a_{\rm crit}$, $R_0$, and $\beta$ are the same as in Equation~\ref{eq:emri_cir_ana}.
At $12$\,Gyr in model M5, the values of $N_{\rm obj,0}$ for SBHs, NSs, WDs, and BDs within $R_0$ are
$\sim 2\times10^3, 1.4\times10^3, 6.4\times10^3$, and $3.5\times10^4$, respectively.
In the case of a Schwarzschild MBH, substituting these values into Equation~\ref{eq:emri_cir_ana},
we find that $\Gamma_{\rm EMRI}$ for SBHs, NSs, WDs, and BDs is $\sim 3\times10^{-8}$\pyr,
$10^{-9}$\pyr, $4\times10^{-9}$\pyr, and $10^{-9}$\pyr, respectively. These rates are consistent within one order
of magnitude with the EMRI rates at $12$\,Gyr shown in Figure~\ref{fig:emris_lc5}.

For plunge events, all these compact objects have relatively high rates
of $10^{-6}\sim 10^{-4}$\pyr, which depend weakly on the spin of the MBH and the
initial density slope of the cluster.

We find that the long-term evolution of EMRI rates is primarily controlled
by the mean density of the cluster ($\propto M_{\rm cl}/r_{\rm eff}^3$),
while an enhanced initial inner density only increases rates temporarily.
Figure~\ref{fig:emris_lc5} shows the EMRI evolution in models M5 and M5\_2,
both of which have the same initial effective size, but the latter
has a higher inner density. We can see that both models have
almost the same EMRI rates for SBHs, NSs, WDs, and BDs after $\sim 1$\,Gyr,
although initially the rates in model M5\_2 are about one order of magnitude higher.

\subsubsection{Including the MBH's mass growth}
\label{subsec:fix_comp_mass_growth}

\begin{figure*}
    \center
\includegraphics[scale=0.8]{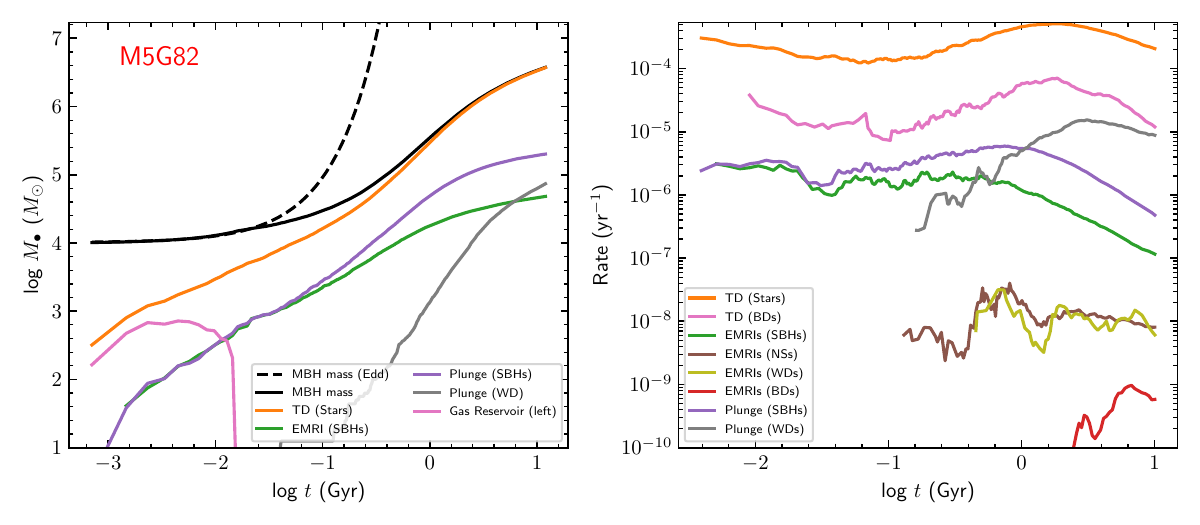}
\caption{
Left panel: Evolution in Model M5G82 (see Table~\ref{tab:modelnoSE}) of the MBH mass (black solid line),
the theoretical mass growth under the Eddington limit (black dashed line), the current
mass of the gas reservoir ("Gas reservoir (left)"), and the cumulative contributions from TDEs
of stars ("TD (stars)"), SBH-EMRIs ("EMRI (SBHs)"), and plunge events of SBHs and WDs ("Plunge (SBHs)" and "Plunge (WDs)").
Contributions from plunge/EMRI events of both NSs and BDs are negligible and
are thus not shown for clarity.
Right panel: Evolution in the same model of the event rates of
TDEs of stars or BDs, EMRIs, and plunge events of compact objects.}
\label{fig:mbh_m5g82}
\end{figure*}
We also explore the evolution when the MBH mass is growing.
The method is similar to that in Section 4.2 of~\citetalias{Paper2}, but here
we additionally include the mass growth of the MBH from swallowing EMRIs.
The simulation results for model M5G82 in Table~\ref{tab:modelnoSE} can be found
in Figure~\ref{fig:mbh_m5g82}, where the MBH mass grows from $10^4\msun$
to $\sim 3.8\times10^6\msun$ over $12$\,Gyr.

The SBH-EMRI event rates gradually decrease from $4\times10^{-6}$\pyr~before
$1$\,Gyr to $\sim 10^{-7}$\pyr~by $12$\,Gyr.
For NSs and WDs, the EMRI rates remain nearly constant at $0.2\sim2\times10^{-8}$\pyr.
For BDs, the rates slowly increase from $10^{-10}$\pyr
at $1$\,Gyr to $\sim 10^{-9}$\pyr~by $12$\,Gyr.
The evolution of EMRI events exhibits different behaviors from those of models with a fixed MBH mass
(M5 or M5\_2 in Figure~\ref{fig:emris_lc5}). Thus, including the mass growth of the MBH is
crucial for accurately modeling the cosmological evolution of EMRI events.

We find that the final MBH mass and the evolution of stellar TDEs and SBH plunge events
in model M5G82 are quite similar to those in model M2G82
from~\citepalias[][see its Figure 10]{Paper2}, although the latter considers
only two mass components (stars and SBHs).


\subsection{NSCs with a mass spectrum, stellar evolution, and stellar mass loss}
\label{subsec:NSCs_sf}
\begin{table*}
\caption{Models with a mass spectrum}
    \centering
\begin{tabular}{|c|c|c|c|c|c|c|c|c|c|c|c|}\hline
    Name & $M_{\rm cl,i}^a$ ($\msun$) & $M_{\rm cl,f}^a$ ($\msun$)& $r_a $
    &$r_{\rm eff,i} $  & $r_{\rm eff,f}$ & Components$^{b}$  & $f_{\rm ma}^c$ & $a^d$ & $M_{\bullet,f}^e$ ($\msun$)
    & $r_{\rm h,f}^e$ \\
\hline
    {MKG91F0k}     & $ 10^9 $ & $5.8\times10^8$ & $2.56$  & $4.66$  & $8.3$       &
    \multirow{11}{*}{\begin{tabular}{c} Kroupa IMF
        \\ MSs ($0.01\msun-150\msun$) \end{tabular}} & $0$ & $1$ & $2.3\times10^7$ &  $2.2$ \\
\cline{1-6}\cline{8-11}
    {MKG91F0.1k}     & $ 10^9 $ & $5.8\times10^8$ & $2.56$  & $4.66$  & $7.7$       &
        & $0.1$ & $1$ & $5.9\times10^7$ &  $3.7$ \\
    \cline{1-6}\cline{8-11}
    {MKG81F0k}  &  $7\times10^7$ & $3.8\times10^7$  &  $0.9$ &  $1.6$  &  $4.9$  &  & $0$   & $1$  & $4.0\times10^6$ & $3.0$ \\
    \cline{1-6}\cline{8-11}
    {MKG82F0k}  &  $7\times10^7$  & $4.0\times10^7$  &  $1.5$ &  $2.7$  &  $6.5$  &  & $0$   & $1$  & $2.0\times10^6$ & $2.5$ \\
    \cline{1-6}\cline{8-11}
    {MKG82F0.06k}  &  $7\times10^7$  & $3.9\times10^7$  &  $1.5$ &  $2.7$  &  $6.8$  &  & $0.06$   & $1$  & $4.0\times10^6$ & $3.9$ \\
    \cline{1-6}\cline{8-11}
    {MKG82F0.06s}  &  $7\times10^7$  & $3.9\times10^7$  &  $1.5$ &  $2.7$  &  $7.0$  &  & $0.06$   & $0$  & $3.9\times10^6$ & $3.9$ \\
    \cline{1-6}\cline{8-11}
    {MKG83F0k}  &  $7\times10^7$ & $4.2\times10^7$ & $2.3$  &  $4.2$ &  $8.4$   &     & $0$ & $1$
    & $8.6\times10^5$ & $2.0$\\
    \cline{1-6}\cline{8-11}
    {MKG83F0.1k}  & $7\times10^7$ & $4.0\times10^7$ & $2.3$  &  $4.2$ &  $9.0$  &   & $0.1$   & $1$ & $4.2\times10^6$ & $5.1$ \\
    \cline{1-6}\cline{8-11}
    {MKG83F1k}  &  $7\times10^7$ & $3.7\times10^7$ & $2.3$  &  $4.2$ &  $5.2$  &      & $1$   & $1$ & $2.6\times10^7$ & $4.5$\\
    \cline{1-6}\cline{8-11}
    {MKG71F0k} & $2\times10^6$ & $1.0\times10^6$& $0.39$  & $0.7$  &  $8.7$ & & $0$ & $1$ & $6.3\times10^4$& $4.3$\\
    \cline{1-6}\cline{8-11}
    {MKG71F0.1k} & $2\times10^6$ & $0.9\times10^6$& $0.39$  & $0.7$  &  $8.6$ & & $0.1$ & $1$ & $1.4\times10^5$& $7.1$\\
    \cline{1-7}\cline{8-11}
    MTG82F0.02k   &  $1.7\times10^8$ & $4.1\times10^7$ & $1.5$   &  $2.7$ &  $20$   &  \multirow{2}{*}{\begin{tabular}{c} Top-Heavy IMF
        \\ MSs ($0.01\msun-150\msun$) \end{tabular}}
    & $0.02$& $1$ & $4.0\times10^6$ & $10$\\
    \cline{1-6}\cline{8-11}
    MTG83F0.025k  &  $1.7\times10^8$ & $4.1\times10^7$ & $2.3$   &  $4.1$ &  $24$   &  & $0.025$& $1$ & $4.0\times10^6$ & $11$\\
    \cline{1-6}\cline{8-11}

    \hline
\end{tabular}
%
    \tablecomments{Models similar to Table~\ref{tab:modelnoSE} but with stellar evolution.
    All models use $\gamma=1$ in Dehnen's model. Initially, the mass of the MBH is $10^4\msun$.
    The mass growth of the MBH is now due to: accreting gas released by TDEs and a fraction $f_{\rm ma}$
    of stellar mass loss, and swallowing EMRIs, plunge events, and any stellar objects falling directly into the loss cone.\\
    $^a$. $M_{\rm cl,i}$ and $M_{\rm cl,f}$ are the initial cluster mass and the mass at $12$\,Gyr, respectively.\\
    $^b$. Kroupa IMF follows~\citet{2001MNRAS.322..231K}; Top-Heavy IMF is similar to Kroupa IMF but with $\alpha=-1.6$
    for $m_\star\in(0.5\msun,150\msun)$.\\
    $^c$. $f_{\rm ma}$ is the fraction of gas from stellar mass loss that is added to the gas reservoir.\\
    $^d$. $a$ is the dimensionless spin parameter of the MBH. \\
    $^e$. $M_{\bullet,f}$ and $r_{\rm h,f}$ are the MBH mass and influence radius at 12\,Gyr, respectively.
    The influence radius is defined by $M_\star(<r_h)=2\bh$, where $M_\star(<r)$ is the enclosed stellar mass
    within $r$.
}    %
\label{tab:modelSE}
%
\end{table*}

In this section, we extend our studies by including three additional complexities in the simulation:
components with a spectrum of masses, stellar evolution processes (see details in Section~\ref{subsec:stellar_evolution}),
and the additional mass growth of the MBH from accreting gas released by mass loss during stellar evolution.
The initial conditions of the models investigated in this section
are shown in Table~\ref{tab:modelSE}.

We consider two types of IMFs. The first follows the
Kroupa IMF~\citep{2001MNRAS.322..231K} for MSs ranging from $0.01\msun$ to $150\msun$.
The number distribution of stars in the Kroupa IMF follows
$f(m_\star)\propto m_\star^{\alpha}$, where $\alpha=-0.3$ if $m_\star\in (0.01,0.08)\msun$,
$\alpha=-1.3$ if $m_\star\in (0.08,0.5)\msun$, and $\alpha=-2.3$ if $m_\star\in (0.5,150)\msun$.
Observations in our Galactic Center suggest that the NSC may follow a top-heavy IMF with $\alpha=-1.6\sim -0.45$
\citep{2010ApJ...708..834B,2013ApJ...764..155L}. Therefore, we implement a second IMF model that is similar to the
Kroupa IMF but adopts $\alpha=-1.6$ for $m_\star\in (0.5,150)\msun$.
For all models, we assume a metallicity of $Z=0.02$.

The diffusion coefficients are calculated by \GNC~in discrete mass bins~\citepalias{Paper2}.
For all models in Table~\ref{tab:modelSE}, we define mass bin edges at
$0.01\msun$, $0.1\msun$, $0.5\msun$, $1\msun$, $2\msun$, $4\msun$,
$8\msun$, $12\msun$, $24\msun$, $32\msun$, $64\msun$, $128\msun$, and $150\msun$, resulting in
$12$ mass bins for the evaluation. We find that this number of mass bins is sufficient for the
convergence of the model results.

A significant amount of mass is expected to be lost during stellar evolution. For example,
when a star evolves from the main sequence
to the giant branch and finally to a compact object, a significant amount of its mass is lost during these transitions.
We assume that a fraction $f_{\rm ma}$ of the lost mass is added to a gas reservoir,
which can be subsequently accreted by the central MBH. Since the majority of the MBH's mass growth is expected to come from
stellar evolution mass loss~\citep[e.g.,][]{2002A&A...394..345F,2006ApJ...649...91F},
the factor $f_{\rm ma}$ plays a crucial role in determining the long-term evolution of
the MBH mass and the structure of the NSC.

There are large uncertainties in $f_{\rm ma}$, as it is possible that some gas
may fragment and form stars before falling into the center and being accreted by the MBH.
\citet{1991ApJ...370...60M} adopt a fraction of $100\%$.
\citet{2006ApJ...649...91F} make an ad hoc assumption of $f_{\rm ma}=6.53\%$ to reproduce $\bh/M_{\rm cl}=0.05$ after $\sim10$\,Gyr.
In this study, we explore $f_{\rm ma}$ ranging from $0$ to $1$, as shown
in Table~\ref{tab:modelSE}.

We primarily focus on models that can reproduce Milky-Way-like NSCs with masses of $\sim 4\times10^7\msun$ by $12$\,Gyr.
These models start with cluster masses of $M_{\rm cl}=7\times10^7\msun$ or $1.7\times10^8\msun$, depending on the
IMF. We also study more massive NSCs with initial masses of $M_{\rm cl}=10^9\msun$ and small NSCs with $M_{\rm cl}=2\times10^6\msun$.

We first provide detailed analyses of the co-evolution of the NSC and the MBH,
including the evolution of stellar components (Section~\ref{subsec:evl_number}),
the NSC structure (Section~\ref{subsec:evl_cluster}), TDEs (Section~\ref{subsec:evl_tde}), and
MBH mass growth (Section~\ref{subsec:evl_mbh}).
These results are critical for understanding the formation and evolution of EMRIs, as together they
establish the environmental background where EMRIs form.
In Section~\ref{subsec:evl_milkyway}, we focus specifically on
models that can reproduce the Milky-Way NSC.

We then analyze the evolution of EMRI events in
Section~\ref{subsec:evl_emri_rate}. The expected mass, eccentricity, and orbital inclination distributions in the LISA band will be discussed
in Section~\ref{subsec:evl_emri_ecc} and~\ref{subsec:evl_emri_inc}, respectively.


\subsubsection{The evolution of stellar populations}
\label{subsec:evl_number}
\begin{figure*}
    \center
\includegraphics[scale=0.6]{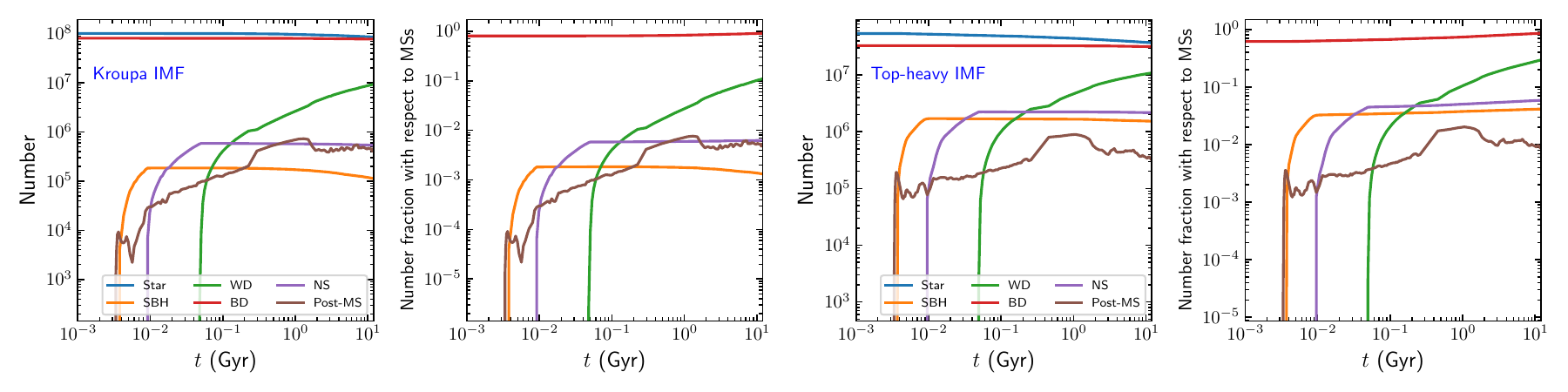}
\caption{First panel: Evolution of the number of different stellar objects assuming a Kroupa IMF,
taken from the results of Model MKG82F0.06k in Table~\ref{tab:modelSE}.
Second panel: Similar to the first panel but for the number ratio
of different stellar objects to MSs. Third and fourth panels:
Similar to the first and second panels but for
a top-heavy IMF, taken from the results of Model MTK82F0.02k.}
\label{fig:nfrac}
\end{figure*}

The numerical evolution of MSs, Post-MSs, BDs, and compact objects plays an important role in
understanding the evolution of NSCs and EMRIs. The evolution of the stellar populations
for the Kroupa IMF and the top-heavy IMF is shown in Figure~\ref{fig:nfrac}.

We can see that for the Kroupa IMF, the number ratios of SBHs, NSs, WDs, BDs, and Post-MSs
to MSs are approximately $1.3\times10^{-3}$, $6.3\times10^{-3}$, $0.11$, $0.91$, and
$5.4\times10^{-3}$ at $12$\,Gyr, respectively. Since SBHs and NSs are remnants of massive MSs ($\gtrsim8\msun$)
with relatively short lifetimes, their populations rapidly peak around
$\sim 0.01-0.1$\,Gyr and then remain nearly constant. In contrast,
the WD population grows continuously over cosmic time due to the evolution
of low-mass stars. The Post-MS population generally increases until $\sim 1$\,Gyr,
after which it exhibits minor variations due to complex post-main-sequence evolution.

If we alternatively adopt a top-heavy IMF with $\alpha=-1.6$ for $m\gtrsim0.5\msun$, the evolution
is similar, but the populations of Post-MSs and compact objects increase significantly.
By $12$\,Gyr, the number ratios of SBHs, NSs, WDs, BDs, and Post-MSs to MSs
become $0.041$, $0.059$, $0.29$, $0.85$, and $8.6\times10^{-3}$, respectively.


\subsubsection{The evolution of the NSC's structure}
\label{subsec:evl_cluster}

\begin{figure*}
    \center
\includegraphics[scale=0.8]{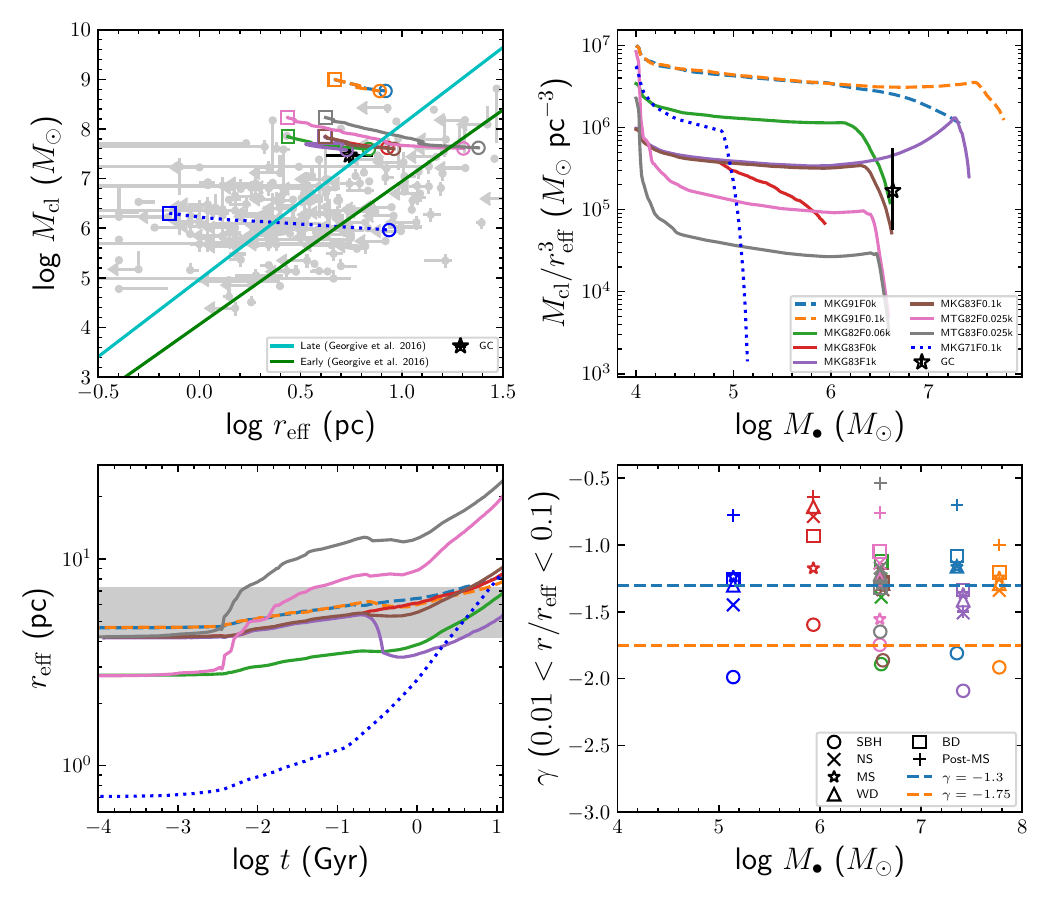}
\caption{Top left panel: Cluster mass $M_{\rm cl}$ and   effective radius $r_{\rm eff}$
of some models in Table~\ref{tab:modelSE}.
For each model, the point on the left represents the initial
value, and the point on the right represents the value at $12$\,Gyr.
The gray circles with error bars indicate observations of nearby galaxies from~\citet{2016MNRAS.457.2122G}.
The cyan and dark-green solid straight lines represent the fitted correlation between
$M_{\rm cl}$ and $r_{\rm eff}$ by~\citet{2016MNRAS.457.2122G} for
late- and early-type galaxies, respectively.
The black empty star marks the value in our Galactic Center.
Top Right panel: The evolution of the mean density of the cluster, $M_{\rm cl}/r_{\rm eff}^3$, as
a function of the mass of the MBH, $\bh$.
Bottom left panel: The evolution of $r_{\rm eff}$ as a function of time.
The shaded region indicates the current constraints on the Milky Way NSC's effective radius
($4.2$\,pc$<r_{\rm eff}<7.2$\,pc)~\citep{2020A&ARv..28....4N}.
Bottom right panel: Density slope indices of different stellar
objects versus the present-day mass of the MBH.
Symbols for each model have the same color
as the lines in the middle panel.
The slope indices are calculated based on the densities
between $10^{-3}r_{\rm eff}<r<0.1r_{\rm eff}$.
    }
\label{fig:nsc_evl}
\end{figure*}

\begin{figure*}
    \center
\includegraphics[scale=0.6]{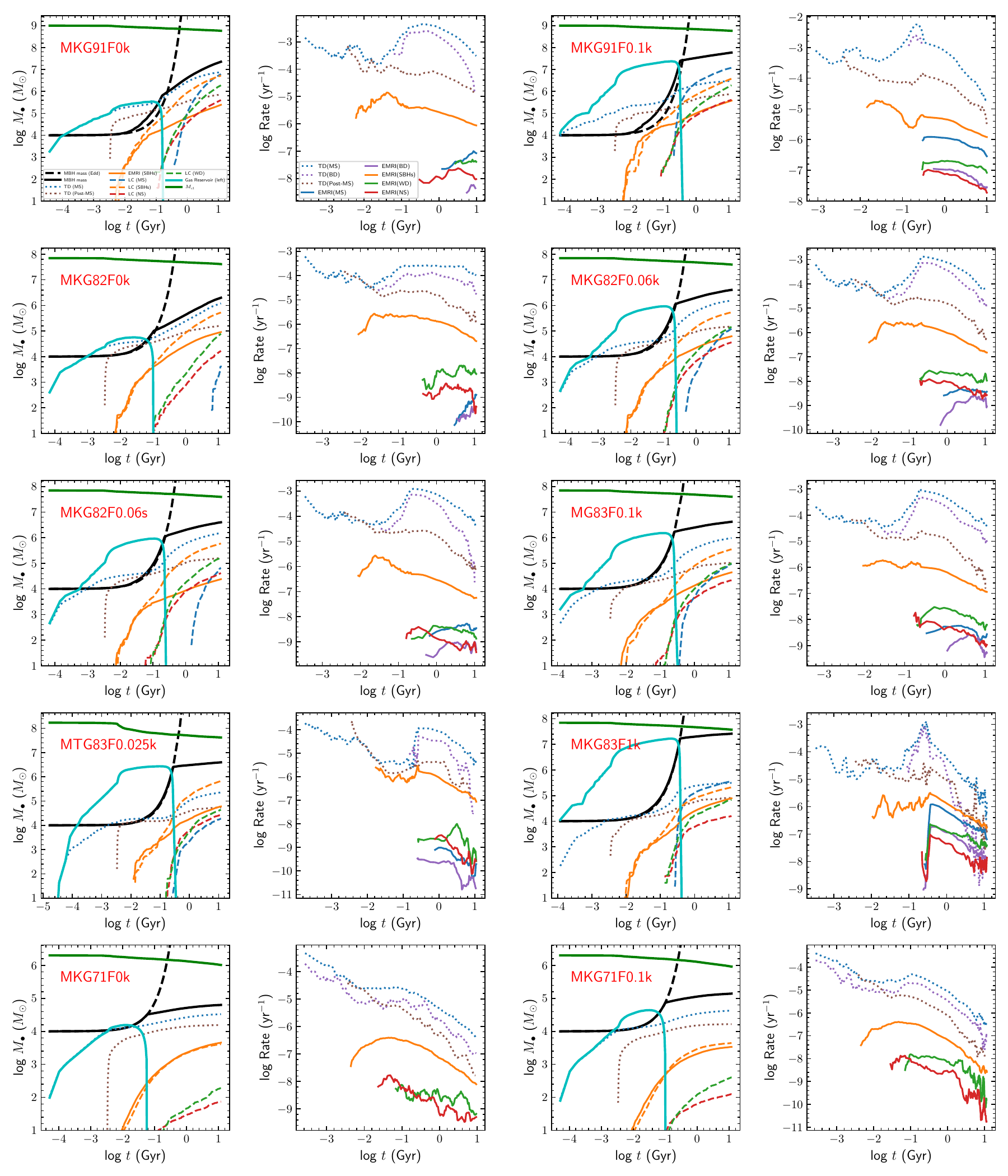}
\caption{The mass growth of the MBH (first and third columns of panels) and the
event rate evolution of EMRIs and TDEs (second and fourth columns of panels)
for different models in Table~\ref{tab:modelSE}. For each model, the legends are similar to those described in Figure~\ref{fig:mbh_m5g82} (additionally, "TD(Post-MS)" means
TDEs of Post-MSs). The event rates for each model have been smoothed
via a moving average over every five or ten data points. }
\label{fig:mbh_evl_dehnen}
\end{figure*}

The evolution of the effective radius ($r_{\rm eff}$), the cluster's stellar mass ($M_{\rm cl}$),
 and the mean density of the cluster ($M_{\rm cl}/r_{\rm eff}^3$)
can be found in Figure~\ref{fig:nsc_evl}.
The MBH mass and the effective size of the cluster at $12$\,Gyr can be found in Table~\ref{tab:modelSE}.

We can see that NSCs exhibit significant size expansion and a decrease in mean density over cosmic time.
The size evolution of an NSC is mainly affected by three factors:
\begin{itemize}
    \item Mass loss from the cluster due to stellar evolution, which
    weakens the stellar gravitational potential and causes stellar orbits to expand.
    As listed in Table~\ref{tab:modelSE}, the cluster loses approximately $40\%$ of its mass over $12$\,Gyr
    for a Kroupa IMF or $\sim 75\%$ for a top-heavy IMF, with most of the mass loss occurring within the
    first $100$\,Myr. The larger the mass loss relative to the NSC's mass, the more pronounced the expansion.
    \item Stellar relaxation processes, which cause the cluster to expand (see~\citet{2009ApJ...694..959M}
    and Section 4.2.1 of~\citetalias{Paper2}). The expansion also facilitates the escape of particles from the system.
    By $12$\,Gyr, we find that the mass that has escaped from the NSC is $<1\%$, $\sim 3\%$, and $\sim 15\%$ of the stellar mass
    for massive NSCs (e.g., model MKG91F0k), Milky-Way-like NSCs (e.g., model MKG83F0k),
    and small NSCs (e.g., model MKG71F0k), respectively.
    \item When the $e$-folding timescale of MBH mass growth--the time required for its mass to increase by a factor of $\sim2.7$, also known as the Salpeter timescale--is much shorter than the local relaxation timescale, the rapid growth of MBH's mass can slow or even reverse the size expansion of the cluster.
    In this case, the conservation of radial action (see Section 2.3 in~\citetalias{Paper2}) causes
    stellar objects to sink toward the center, increasing the cluster's density and reducing its effective size
    (this effect is more apparent for model MKG83F1k in the bottom left panel of Figure~\ref{fig:nsc_evl}).
\end{itemize}

Thus, the typical evolution of an NSC's size proceeds as follows:
In early epochs ($<100$\,Myr), stellar mass loss accelerates the size expansion.
When the gas reservoir is large enough to sustain Eddington-limited
accretion, the MBH's mass growth is fast, slowing down or even reversing the expansion process.
However, once the MBH mass is so large that the gas reservoir can no
longer sustain Eddington-limited accretion, its growth rate declines, and
stellar relaxation processes become dominant. This transition ultimately leads to cluster expansion.

At $12$\,Gyr, the density slope of SBHs, averaged in the region
$r/r_{\rm eff}\in(0.01,0.1)$, is $\sim -2.0$ (bottom right panel of Figure~\ref{fig:nsc_evl}).
In contrast, lighter objects, e.g., MSs, NSs, WDs, and BDs, exhibit a much shallower density profile, with
slope indices ranging from $-1.5$ to $-0.5$.
These results are found to be insensitive to the initial conditions of the cluster.
Thus, the present-day density profile
of a cluster with a spectrum of masses remains similar to the cases
where the cluster contains only two components (stars and SBHs), as explored
in~\citetalias{Paper1}.

\subsubsection{The evolution of TDE rates}
\label{subsec:evl_tde}
It is necessary to investigate the evolution of TDEs, as they can contribute a non-negligible
fraction of mass to the cosmological mass growth of MBHs.
Figure~\ref{fig:mbh_evl_dehnen} shows the cosmological evolution of TDEs of MSs, BDs, and Post-MSs in different models
(dotted lines in the panel). Similar to~\citetalias{Paper2}, we find that the TDE rates of both MSs and BDs
evolve depending on the ratio between the
gravitational influence radius of the MBH, $r_h$ (defined by $M_\star(<r_h)=2\bh$, where $M_\star(<r)$
is the enclosed stellar mass within $r$), and the cluster's effective radius, $r_{\rm eff}$.
If initially $r_h\lesssim 0.08r_{\rm eff}$, the rates evolve in three distinct phases: (1) an initially high rate
that decreases with time; (2) a subsequent rise driven by the rapid, Eddington-limited mass growth of the MBH;
and (3) a decline again once the MBH mass is sufficiently large that $r_h>0.1r_{\rm eff}$.
In contrast, if initially $r_h\gtrsim 0.08r_{\rm eff}$, the evolution skips the first two phases and
proceeds directly to phase 3.

The bottom right panel of Figure~\ref{fig:m_prop} shows
 the TDE rates of MSs averaged over the early universe ($<1$\,Gyr) and later epochs ($1\sim12$\,Gyr).
There is a systematic decrease in TDE rates over cosmic time.

For Post-MSs, they do not follow the above behavior, as their TDE rates
also depend on their evolving population size (see Figure~\ref{fig:nfrac})
and the evolving distribution of their stellar radii. When including these complexities,
Figure~\ref{fig:mbh_evl_dehnen} shows that their TDE rates generally exhibit an initial peak and then
gradually decline over time.

The TDE rates of MSs are generally higher than those of BDs,
which are in turn higher than those of Post-MSs. The TDE rates of BDs and Post-MSs are
one or two orders of magnitude lower.

In all models, the peak TDE rates of
MSs are usually $10^{-3}\sim 10^{-2}$\pyr~for massive NSCs ($M_{\rm cl} = 10^9\msun$) and
$10^{-4}\sim 10^{-3}$\pyr~for Milky-Way-like or smaller NSCs ($M_{\rm cl} = 2\times10^6\msun$). However,
at $12$\,Gyr, the rates reduce to $10^{-4}\sim 10^{-3}$\pyr~ for massive NSCs,
$10^{-5}\sim 10^{-4}$\pyr~for Milky-Way-like NSCs, and $10^{-7}\sim 10^{-6}$\pyr~
for smaller NSCs. These results are generally consistent with~\citetalias{Paper2}.
However, while \citetalias{Paper2} assumed equal-mass MSs of $1\msun$, we find here that
the majority of MS-TDEs involve stars with masses of $\sim0.5\msun$, which is
two times smaller.

\subsubsection{Mass growth of the MBH}
\label{subsec:evl_mbh}

The mass growth of the MBH over $12$\,Gyr in different models can be found in Figure~\ref{fig:mbh_evl_dehnen}.
In the early Universe, MBHs typically experience a phase of rapid mass growth through
Eddington-limited accretion, lasting approximately $0.1-0.5$\,Gyr. This phase is
mainly sustained by gas supplied from TDEs
and stellar mass loss (if included). The duration of this phase depends on
both the MBH's gas consumption rate (which is higher for a more massive MBH)
and the gas supply rate.

The mass increase of the MBH at $12$\,Gyr is approximately the sum of $\delta M_{\bullet, \rm td}$
(mass contributed by accreting gas from TDEs), {$\delta M_{\bullet, \rm d}$ (mass contributed 
by direct swallowing of stellar objects)} and $\delta M_{\bullet, \rm ml}$ (mass contributed by stellar mass loss).
Then the final mass of the MBH at $12$\,Gyr is
\be\ba
&M_{\bullet,f}  \simeq\delta M_{\bullet, \rm td}+\delta M_{\bullet, \rm d}+\delta M_{\bullet, \rm ml}+M_{\bullet,i}\\
   \simeq & \langle m_\star\rangle_{\rm td} \bar R_{\rm td,MS} 
   T_{12\rm\,Gyr}+\langle m \rangle_{\rm d} \bar R_{\rm d} T_{12\rm\,Gyr}\\
   +&M_{\rm cl,i}f_{\rm ml} f_{\rm ma}+M_{\bullet,i}
\ea\ee
where
\begin{itemize}
\item $\langle m_\star\rangle_{\rm td}\sim 0.5\msun$ is the mean mass of TDEs (see Section~\ref{subsec:evl_tde});
\item $\bar R_{\rm td,MS}$ is the mean event rate of MS-TDEs;
\item $T_{12\rm\,Gyr}=12$\,Gyr is the time of {evolution};
\item {$\langle m \rangle_{\rm d}$ and $\bar R_{\rm d}$ is the mean mass and event rate of objects directly swallowed by the central MBH};
\item  $M_{\rm cl,i}$ is the initial mass of the cluster;
\item $f_{\rm ml}=0.4\sim0.7$ (see Section~\ref{subsec:evl_cluster}) is the fraction of stellar mass lost
relative to the initial cluster mass;
\item $f_{\rm ma}$ is the assumed fraction of mass loss that is added to the gas reservoir, which can
be further accreted by the MBH.
\item $M_{\bullet,i}=10^4\msun$ is the initial MBH mass ($M_{\bullet,i}=10^4\msun$).
\end{itemize}

According to Section~\ref{subsec:evl_tde} and Figure~\ref{fig:mbh_evl_dehnen},
the mean MS-TDE rates are generally
$\bar R_{\rm td,MS}\sim 10^{-3}$\pyr, $\sim 10^{-4}$\pyr, and $\sim 10^{-6}$\pyr~
for massive NSCs ($M_{\rm cl,i}=10^9\msun$), Milky-Way-like NSCs, and smaller
NSCs ($M_{\rm cl,i}=2\times10^6\msun$), respectively.
The total mass contributed by TDEs over $12$\,Gyr is
typically $\delta M_{\bullet, \rm td} \sim 10^7\msun$, $\sim 10^6\msun$, and $\sim 5\times10^4\msun$
for massive, Milky-Way-like, and smaller NSCs, respectively.
These results are consistent with the final MBH mass for models with $f_{\rm ma}=0$,
e.g., models MKG91F0k, MKG82F0k, and MKG71F0k.

Note that since $\langle m_\star\rangle_{\rm td}\sim 0.5\msun$,
the mass contributed by TDEs is about two times smaller compared to the models
of~\citetalias{Paper2}, which assume MSs of $1\msun$.

{The total rate of objects falling into the loss cone is given by 
$\bar {R}_{\rm lc}\simeq \bar R_{\rm td, MS}+\bar R_{\rm d}$. 
The contribution from the direct swallowing of objects is usually smaller than that from 
TDEs. However, it can become significant and even larger than those from TDEs
for MBHs with masses $\bh\gtrsim 10^7\msun$. For example, in model MKG91F0k,
which has a final MBH's mass of $\bh\sim 2\times10^7\msun$, the contribution from 
direct capture can reach up to $\sim 60\%$.
}

Similar to~\citetalias{Paper2}, we find that the smaller the initial effective radius
($r_{\rm eff,i}$) of the NSC, the larger the rate of MS-TDEs ($R_{\rm td,MS}$),
and thus the larger the final MBH mass at $12$\,Gyr.

The mass contributed by stellar mass loss, $\delta M_{\bullet, \rm ml}$,
depends critically on the assumed value of $f_{\rm ma}$.
From Table~\ref{tab:modelSE}, we can see that for models assuming $f_{\rm ma}\ge0.1$,
the final MBH mass is dominated by $\delta M_{\bullet, \rm ml}$, and the MBH-to-NSC
mass ratio can be up to $\sim 0.1$ ($\sim 0.8$) if we assume $f_{\rm ma}=0.1$ ($f_{\rm ma}=1$).

\subsubsection{Reproducing the Milky Way's NSC}
\label{subsec:evl_milkyway}

\begin{figure*}
    \center
\includegraphics[scale=0.7]{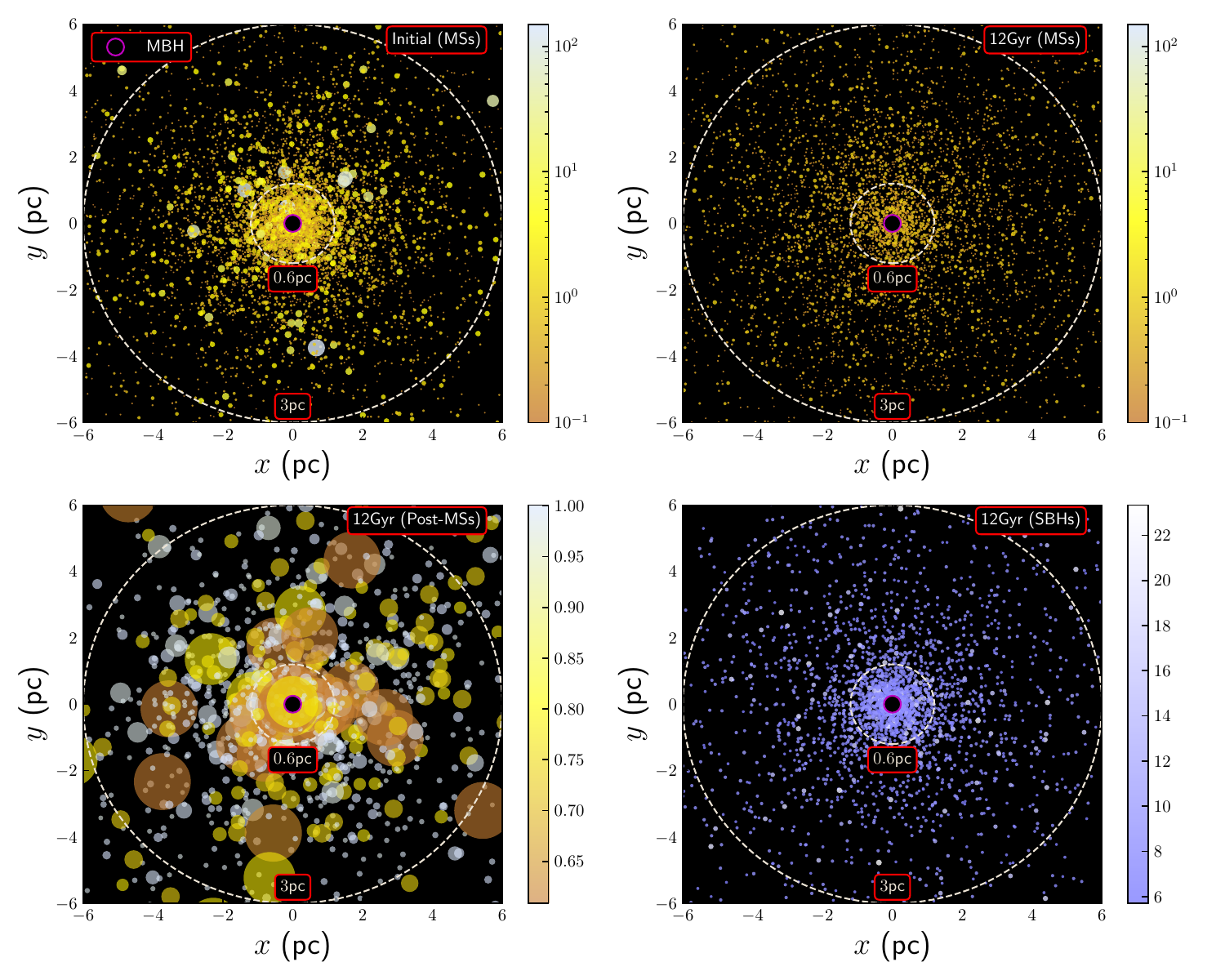}
\caption{The spatial distribution of stellar objects illustrated based on model MKG82F0.06k.
Stellar objects are assumed to be spherically distributed. For clarity, the number of
MSs, Post-MSs, and SBHs is suppressed by a factor of $10^4$, $500$, and $50$, respectively.
The top left panel shows the initial distribution of MSs (all zero-age MSs; BDs
are not plotted for clarity), while the
rest of the panels show the distribution of MSs, Post-MSs, and SBHs at $12$\,Gyr.
In all panels, filled colored circles represent individual stellar objects, with
their sizes proportional to the stellar radii and their color representing the
stellar mass according to the color bar on the right of each panel (in units of solar masses).
The largest MS and Post-MS have
radii of $20R_\odot$ and $226R_\odot$, respectively.
}
\label{fig:ill_mkg83}
\end{figure*}

\begin{figure*}
    \center
\includegraphics[scale=0.6]{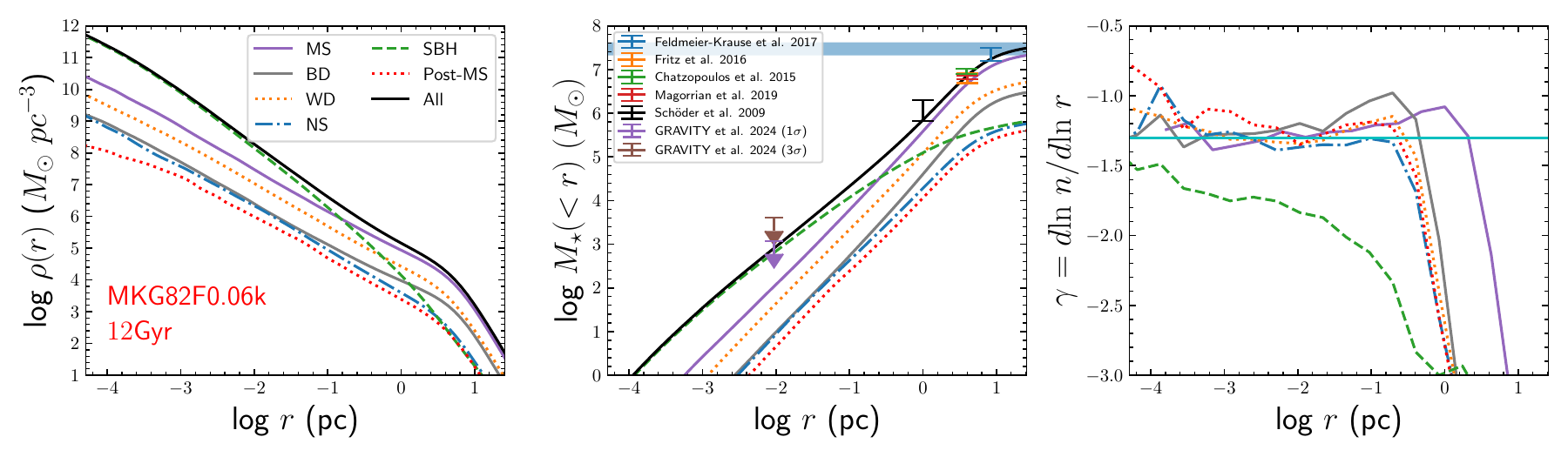}
\caption{Simulation results for different stellar objects at $12$\,Gyr for
model MKG82F0.06k. Left panel: The mass density distribution as a function of radius.
Middle panel: The cumulative stellar mass distribution as a function of radius.
The data with error bars
represents the constraints observed in the Milky Way's NSC, while
the shaded region marks the observational constraints on the total mass
of the Milky Way's NSC ($2.1\sim4.2\times10^7\msun$)
\citep{2009A&A...502...91S,2014A&A...566A..47S,2014A&A...570A...2F,2015MNRAS.447..948C,
2016ApJ...821...44F,2017MNRAS.466.4040F,2019MNRAS.484.1166M, 2024A&A...692A.242G}.
Right panel: The slope index of the density as a function of radius.
The gray solid line shows a reference slope index of $\gamma=-1.3$.
}
\label{fig:ge_mkg83}
\end{figure*}

It is of particular interest to investigate models that reproduce the Milky Way's NSC, for which
we have the best observational constraints.
From Table~\ref{tab:modelSE}, we can
see that models MKG81F0k and MKG82F0.06k(s) can reproduce the current
mass of the MBH in the Galactic Center, i.e., $4\times10^6\msun$~\citep{2009ApJ...692.1075G,2017ApJ...837...30G},
and have an effective size in the range of $4.2$\,pc$\sim7.2$\,pc~\citep{2020A&ARv..28....4N}.

As discussed in Section~\ref{subsec:evl_mbh}, the mass growth of the MBH is contributed
mainly by both TDEs of MSs (which can be regulated by
the initial effective size of the cluster, $r_{\rm eff,i}$) and
mass loss from stellar evolution (which depends critically
on the assumed value of $f_{\rm ma}$).
Model MKG81F0k ignores stellar mass loss ($f_{\rm ma}=0$) and thus requires a compact initial cluster size
($r_{\rm eff,i}=1.6$\,pc) to reproduce the observed MBH mass at $12$\,Gyr solely via TDEs.
Model MKG82F0.06k includes stellar mass loss with an assumed value of $f_{\rm ma}=0.06$
and thus can initially have a slightly larger cluster size.

Note that to reproduce a Milky-Way-like NSC, $f_{\rm ma}$ cannot be larger than $\sim 0.1$. Otherwise,
either the MBH mass or the effective size of the cluster would be larger than the observed values.
This indicates that most of the gas released by stellar mass loss may not be efficiently
accreted by the MBH. One possibility is that the gas collapses and forms stars
before approaching the vicinity of the MBH.

Figure~\ref{fig:ill_mkg83} illustrates the projected
spatial distribution (assuming a spherical distribution), physical size, and masses of stellar objects
for Model MKG82F0.06k.
The initial cluster consists of zero-age MSs with masses ranging from $0.01-150\msun$. After $12$\,Gyr, the MSs remaining in the cluster are those with masses
$\lesssim 1\msun$, as more massive stars have already evolved into post-MSs or compact objects.
At $12$\,Gyr, the post-MSs are in the mass range of $0.6\sim 1\msun$, with much larger sizes of $1.5-230R_\odot$.
From Figure~\ref{fig:ill_mkg83}, it is apparent that SBHs are more concentrated in the inner regions
than MSs, mainly due to mass segregation effects.

The present-day density profile for model MKG82F0.06k is shown in Figure~\ref{fig:ge_mkg83}.
This model can approximately reproduce the observed enclosed mass distribution (middle panel).
For lighter components (MSs, Post-MSs, NSs, WDs, and BDs), they all have
slope indices of $\sim -1.3$ within $\sim 0.1$\,pc. The SBH population has a steeper slope index of $-1.5\sim -2.5$.
These results are roughly consistent with model M2G82 in~\citetalias{Paper2}, although the
latter adopts a simple two-component model and lacks stellar evolution.

\subsubsection{The evolution of EMRI event rates}
\label{subsec:evl_emri_rate}

\begin{figure*}
    \center
\includegraphics[scale=0.75]{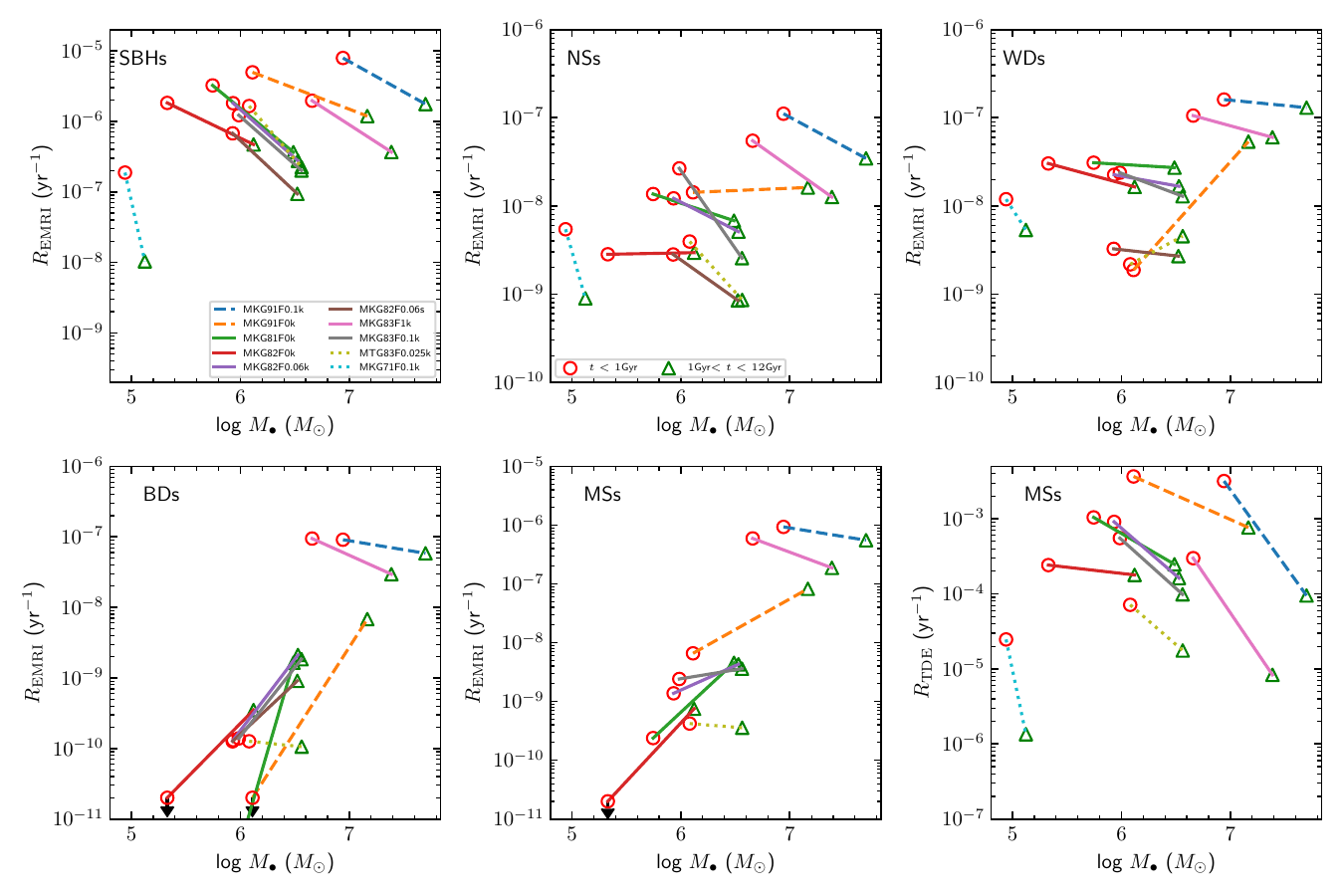}
\caption{The event rates of EMRIs ($R_{\rm EMRI}$) or TDEs ($R_{\rm TDE}$) averaged over the early universe
($t<1$\,Gyr, empty red circles)
and later epochs ($1$\,Gyr$<t<12$\,Gyr, empty green triangles) for some models
in Table~\ref{tab:modelSE}. In the bottom middle and right panels, the results of
Model MKG82F0.06s are similar to those of Model MKG82F0.06k and
are therefore not shown for clarity.
}
\label{fig:m_prop}
\end{figure*}

The evolution of EMRI event rates over $12$\,Gyr can be found in Figure~\ref{fig:mbh_evl_dehnen}.
For a clearer comparison of rates versus MBH masses across different models and cosmic times (
averaged over early epochs, the first $1$\,Gyr, and later epochs, between $1-12$\,Gyr), see Figure~\ref{fig:m_prop}.

According to these results, we find that the evolution of EMRI rates is mainly affected by the following four factors:
\begin{itemize}
    \item The size expansion or contraction of the cluster (see also Section~\ref{subsec:evl_cluster}).
       The expansion of the cluster's size due to
       either substantial stellar mass loss or relaxation processes leads to a decrease in the cluster's density. As
       $R_{\rm EMRI}\propto M_{\rm cl}/r_{\rm eff}^3$ approximately, the EMRI rates for all stellar objects
       are expected to decline.
       However, if the growth of the MBH mass is sufficiently fast such that the local relaxation time is
       longer than the $e$-folding time of the MBH's mass ($\sim 4.5\times10^7$~yr), the shrinking of stellar orbits can slow down
       or even reverse the size expansion. If the latter occurs, the EMRI event rates can increase. This effect is
       particularly apparent in model MKG83F1k, where all EMRI rates show a very substantial increase around $\sim 0.5$\,Gyr.
    \item The evolution of the stellar population (see also Section~\ref{subsec:evl_number}).
    Almost all SBHs and NSs form in the early universe, and afterward, the sizes of their
    populations remain almost constant over cosmic time.
    Thus, their rates are affected mainly by the dynamical evolution of the cluster and the MBH.
    However, the rate evolution of WD-EMRIs is slightly complicated by the growing number of their populations over cosmic time (See Figure~\ref{fig:nfrac}).
    \item The IMF of the cluster. The IMF affects the evolution of the number and mass distribution of stellar populations over cosmic time
    (see Section~\ref{subsec:evl_number}). Different IMFs lead to different amounts of stellar mass loss, which affects
    the size expansion of the cluster and the mass growth of the MBH. These complications ultimately lead to differences in
    the observed rates and mass distribution of EMRIs.
    \item The spin of the MBH. Around a maximally spinning MBH, the EMRI event rates of compact objects
    are usually $3\sim 4$ times higher than those around a non-spinning MBH. For MSs and BDs, however, the enhancement is
    less significant, as their loss cone sizes cannot be smaller than their tidal radii.
    \item For BDs and MSs, they become EMRIs only if $\bh\gtrsim 10^5\msun$
    (see Section~\ref{subsec:critic_a}). The ISO pericenters of BDs and MSs are outside their
    tidal radii only if $\bh\gtrsim 10^6\msun$ (see Figure~\ref{fig:bd_mr}).
    Around smaller MBHs, they are more likely to be tidally disrupted when passing through the loss cone. Thus, starting from a seed
    MBH of $10^4\msun$, BD-EMRIs and MS-EMRIs appear at later epochs of the Universe.
    BD-EMRIs and MS-EMRIs form more easily around more massive MBHs, as their tidal radii
    become relatively smaller compared to the ISO. However, how their rates evolve with time is
    also affected by the factors listed above.
\end{itemize}

Combining all the above complexities, we summarize and discuss the detailed evolution of individual 
stellar populations as follows:
\begin{itemize}
    \item SBH-EMRI events first appear at $10$\,Myr and peak in the early
    universe ($\lesssim 0.1$\,Gyr) with rates of $10^{-7}$\pyr~$\sim 10^{-5}$\pyr.
    After that, they gradually decrease by a factor of $5-10$
    by $\sim 12$\,Gyr.

    The evolution of these rates over cosmic time provides a crucial context for interpreting analytical steady-state models focused on the present epoch. For instance, \cite{AmaroSeoaneMonoOligo} analytically investigated the population of Early-EMRIs (E-EMRIs)—sources in the long-lived early inspiral phase. They derived merger rates for the Galactic Center ranging from $10^{-6}$ to $10^{-5}$ yr$^{-1}$ for $10\msun$ SBHs, assuming the system is currently relaxed.

    Our simulations for Milky-Way-like NSCs (e.g., models MKG81F0k and MKG82F0.06k) show peak SBH-EMRI rates of $\sim 3\times 10^{-6}$\pyr~in the early universe ($\lesssim 1$ Gyr), which aligns well with the lower end of these analytical estimates. The simulations show a subsequent decline to $\sim 10^{-7}$\pyr~by 12 Gyr, driven by long-term effects such as cluster expansion and the depletion of SBHs.

    The key concept in \cite{AmaroSeoaneMonoOligo} is the steady-state number of sources currently observable in the LISA band, calculated by multiplying the merger rate by the long residence time ($T_{\rm res} \approx 1.85\times10^5$ years for a $10\msun$ SBH with SNR $> 10$). While our simulations track the long-term cosmological evolution, the analytical steady-state calculation provides an estimate for the present snapshot in time, relevant for LISA observations in the near future. The GNC simulations confirm the underlying dynamical processes—two-body relaxation feeding the GW-driven inspiral (Figure 2). If the current state of the Galactic Center supports the analytically derived merger rates (which are within an order of magnitude of our peak simulated rates), the prediction of a significant population of E-EMRIs (10-25 sources) currently in the band~\citep{AmaroSeoaneMonoOligo} remains a plausible estimate for present-day observations, as the astrophysical conditions are not expected to change significantly over the residence timescale ($10^5$ years).

    \item NS- and WD-EMRIs first appear slightly later than SBH-EMRIs. Their event rates
     initially peak at $\sim 0.1-1$\,Gyr and then gradually decline over cosmic time.
     The decrease in WD-EMRI rates is not as significant as that of NS-EMRIs, as the number of WDs grows with time.
     Typically, the peaks of NS-EMRI (WD-EMRI) event rates are in the range of $10^{-9}\sim 10^{-7}$\pyr
     ($10^{-8}\sim 10^{-7}$\pyr).
    The event rates of both NS- and WD-EMRIs are about $2-3$ orders of magnitude lower than those of SBH-EMRIs.
    \item For MS- and BD-EMRIs, they only appear when the MBH mass is larger than $\sim10^5\msun$. After that, most models
    exhibit an increase in rates with MBH mass. 
    In general, BD-EMRI and MS-EMRI event rates increase from $10^{-11}\sim 10^{-9}$\pyr~for MBHs with
    $\sim 10^{6}\msun$ to $10^{-8}\sim 10^{-7}$\pyr~ for MBHs with $\sim 10^{7}\msun$.
    In most cases, MS-EMRI event rates are slightly higher than those of BD-EMRIs,
    mainly because MSs are more abundant and have larger masses. Models MKG83F1k and MKG91F1k 
    grow their MBH mass quickly to $>10^7\msun$, and thus their rates peak in the early universe. When the phase of rapid mass growth ends, size expansion starts to dominate, and the rates decrease with time.

BD-EMRIs has been specifically investigated analytically by \cite{2019PhRvD..99l3025A}, where they are termed X-MRIs due to the extreme mass ratio ($\sim 10^8$). Our simulations confirm the premise of that work: BDs around Milky-Way-like MBHs can indeed survive tidal disruption (as $r_{\rm td} < r_{\rm ISO}$, see Figure 1) and successfully form X-MRIs.

\end{itemize}

The rates of SBH-EMRIs are orders of magnitude larger than those of other stellar types,
which can be understood from the analysis of the critical radius $a_{\rm crit}$
discussed in Section~\ref{subsec:critic_a}.

For the models that can reproduce the Milky Way's NSC, i.e., models MKG81F0k and MKG82F0.06k,
the EMRI rates of SBHs, NSs, WDs, BDs, and MSs are approximately
$10^{-7}$\pyr, $5\times10^{-9}$\pyr,  $2\times10^{-8}$\pyr, $10^{-9}$\pyr,
and $4\times10^{-9}$\pyr~by $12$\,Gyr, respectively.

\cite{2019PhRvD..99l3025A} predicted a merger rate of 
$\dot{\Gamma}_{\rm X-MRI}\sim 10^{-6}\,\textrm{yr}^{-1}$ 
for the Galactic Center, assuming $\beta=1.5$ for BDs and $\gamma=1.75$ for SBHs. 
Our simulations for Milky-Way-like models (e.g., MKG82F0.06k) yield BD-EMRI rates of 
$\sim 10^{-9}$\pyr~at 12 Gyr. This discrepancy arises for three reasons: 
(1)~\cite{2019PhRvD..99l3025A} assumes a steady solution and no MBH mass growth,
so it is more appropriate to compare the results with the initial rates of 
model M5\_2 in Section~\ref{subsec:fix_mass_emri_rate}. 
As shown in the bottom panel of Figure~\ref{fig:emris_lc5}, the initial BD-EMRI rate is
around $\sim 10^{-7}$\pyr, consistent with the results of \cite{2019PhRvD..99l3025A} within 
one order of magnitude. (2) In \GNC, the rates are time-dependent, and are mainly affected by the 
cluster expansion, the mass growth of MBH and the assumed IMF (see also ection 2.6). The expansion 
of the cluster reduces the central density of BDs, leading to lower rates of EMRIs.
(3) By $12$\,Gyr, the density slope  of BDs in \GNC~is $\beta\sim 1.3$ (See right panel of 
Figure~\ref{fig:ge_mkg83}), which is flatter than the assumed slope of $1.5$. 
A flatter density profile results in 
smaller rates of EMRIs (See Equation~\ref{eq:emri_rate_ana}).

{Thus, the discrepancies are mainly ascribed to the differences between the cluster 
conditions assumed in the steady-state solution and those obtained from our numerical simulation. 
Nevertheless, our numerical simulations validate the dynamical pathway for forming X-MRIs. 
Considering that our simulation results are still subject to several uncertainties, e.g., 
in the initial cluster's conditions and in other complexities not captured in our simulation (see Section~\ref{sec:discuss}), 
we do not intend to exclude these steady-state solutions.
}

The model with a top-heavy IMF (MTG83F0.025k) has very substantial mass loss ($\sim 75\%$ of the total stellar mass), and thus
the size expansion is quite significant (see bottom left panel of Figure~\ref{fig:nsc_evl}). However, it also produces more SBHs and NSs
over cosmic time compared to the case adopting a Kroupa IMF.
Combining these two factors, the rates of SBH-EMRIs and NS-EMRIs are comparable to those in the Kroupa IMF case,
while the rates of EMRIs involving other stellar objects are lower.

\subsubsection{The mass and eccentricity distribution of EMRIs in the LISA band}
\label{subsec:evl_emri_ecc}

\begin{figure*}
    \center
\includegraphics[scale=0.6]{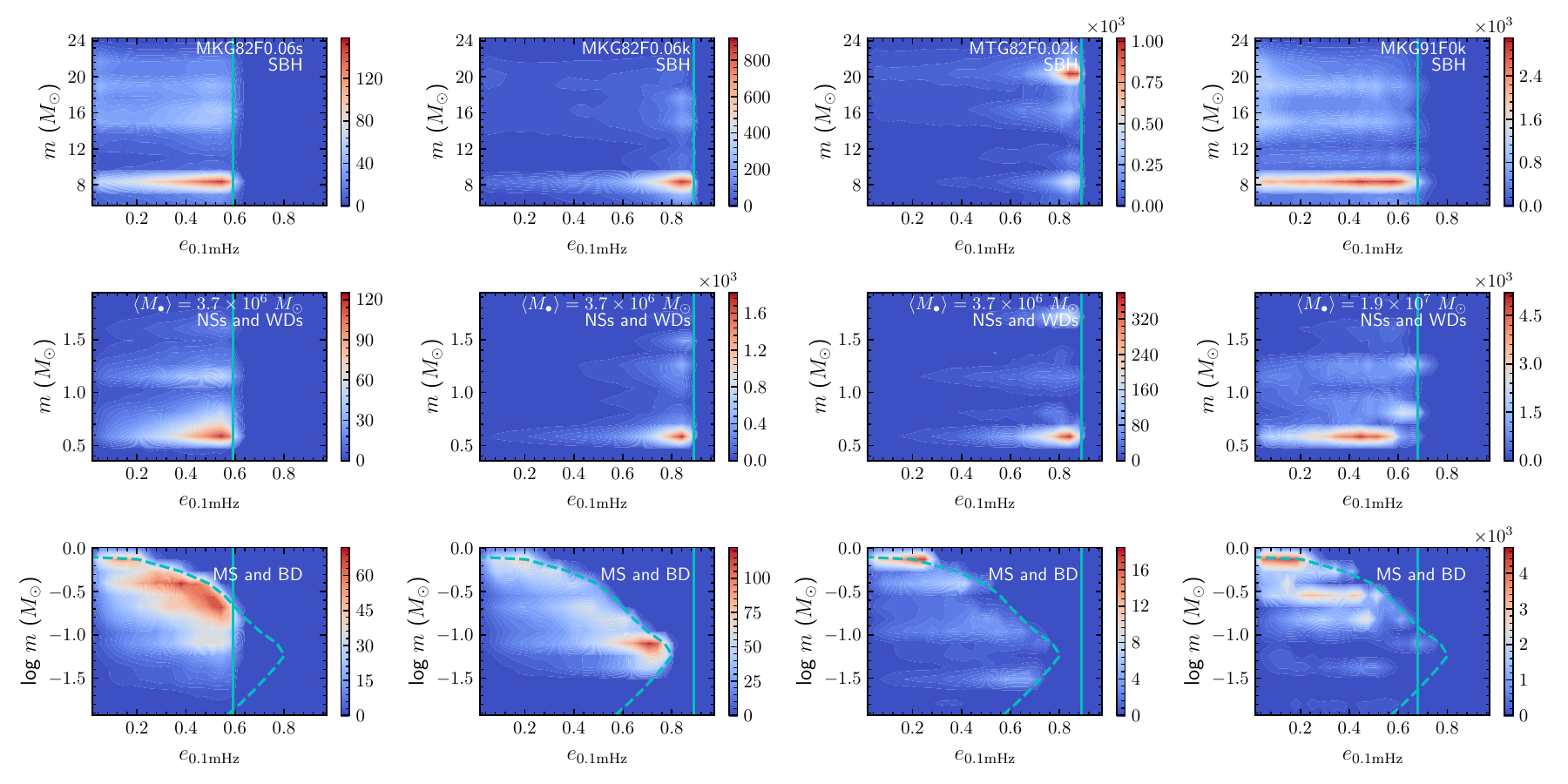}
\caption{The mass-eccentricity joint distribution of EMRI events in Models MKG82F0.06s, MKG82F0.06k,
 MTG83F0.025k, and MKG91F0k.
The distribution is averaged over $5$\,Gyr$\le t\le 12$\,Gyr.
 $\langle \bh\rangle$ shows the mean mass of the MBH during this period.
$e_{0.1\rm mHz}$ is the eccentricity of EMRIs when their orbital periods are $10^4$\,s.
The color contour represents the number of samples per unit of $m$ (or $\log m$) and
$e_{0.1\rm mHz}$.
Panels from top to bottom show the results for SBH-EMRIs, NS- and WD-EMRIs, and MS- and BD-EMRIs, respectively.
The solid cyan lines represent $e_{\rm 0.1mHz,max}$
(Equation~\ref{eq:emax}) for $r_{\rm ISO}(i=0^\circ)=2.12r_{\rm g}$ if the model adopts $a=1$, and
$r_{\rm ISO}=8r_g$ if it adopts $a=0$. In the bottom panels, the dashed cyan lines
show $e_{\rm 0.1mHz,max}$ (Equation~\ref{eq:emax}) based on the tidal radii of MSs or BDs.
}
\label{fig:emris_me}
\end{figure*}

\begin{figure*}
    \center
\includegraphics[scale=0.8]{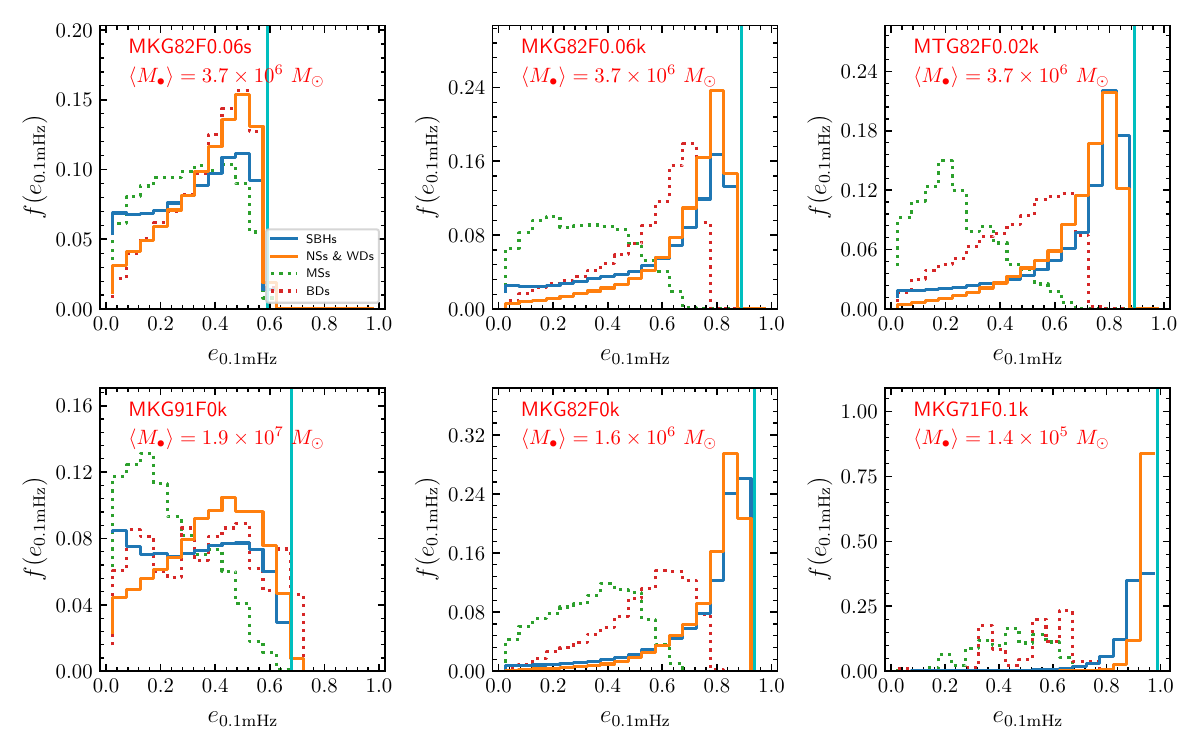}
\caption{The normalized eccentricity distribution of EMRIs for some models in Table~\ref{tab:modelSE}.
Lines in different colors are results for different stellar objects.
The solid cyan vertical line in each panel shows the position of $e_{\rm 0.1mHz,max}$
(Equation~\ref{eq:emax}) for each model.}
\label{fig:emris_eh}
\end{figure*}

Here we investigate the present-day (at $12$\,Gyr in the simulation)
mass and orbital eccentricity distributions of EMRIs in the LISA band.
We focus on the eccentricity of EMRIs ($e_{\rm 0.1mHz}$) when their orbital frequencies
have decayed to $f_{\rm orb}=0.1$\,mHz ($P=10^4$\,s). Their primary GW emissions
are then {$\sim 0.2$\,mHz for circular orbits (or $\sim 5$mHz if $e=0.9)$}, which is within LISA's sensitivity range.

The orbital semi-major axis of these EMRIs {for a given period ($P$)} is
\be
a_{P}\sim18.45~{r_g} \times\left(\frac{P}{10^{4}~{\rm s}}\right)^{2/3}\left(\frac{\bh}{4\times10^6~\msun}\right)^{-2/3}.
\ee
The maximum value of the eccentricity ($e_{\rm 0.1mHz,max}$) is limited by the size of the loss cone:
\be
1-e_{\rm 0.1mHz,max}=0.434\times\frac{r_{\rm p,lc}}{8\,{\rm rg}}\left(\frac{\bh}{4\times10^6~\msun}\right)^{2/3},
\label{eq:emax}
\ee
where $r_{\rm p, lc}=\max(r_{\rm td}, r_{\rm ISO})$.

For a maximally spinning (non-spinning) Milky-Way-like MBH, the highest eccentricity for compact objects
can be up to $\sim 0.88$ ($\sim 0.56$). For BDs and MSs, $e_{\rm 0.1mHz,max}$ is determined not
only by the ISO but also by their tidal radii (see bottom panels of Figure~\ref{fig:emris_me}).
If the loss cone is determined by their tidal radii,
$e_{\rm 0.1mHz,max}\sim0.81$, which is independent of $\bh$.

Note that for an MBH with $a=1$ ($a=0$) and $\bh\gtrsim 10^8\msun$
($\gtrsim 1.4\times10^7\msun$), all EMRIs will have already fallen into the MBH
before their orbits decay to $f_{\rm orb}=0.1$\,mHz.
Similarly, for stellar objects in retrograde orbits, they cannot become EMRIs at $0.1$\,mHz if they are around
maximally spinning MBHs with $\bh\gtrsim 8\times10^6\msun$.

Figure~\ref{fig:emris_me} and~\ref{fig:emris_eh}
show the mass-eccentricity joint distribution and the
orbital eccentricity distribution of EMRIs for some
models in Table~\ref{tab:modelSE}, respectively.

For SBH-EMRIs, their masses are in the range of $5-23\msun$.
The majority of them are around $8\msun$ if assuming
a Kroupa IMF, or $\sim 20\msun$ if assuming a top-heavy IMF.
For Milky-Way-like or smaller MBHs, SBH-EMRIs in the LISA band are
more likely to be found with eccentricities near $e_{\rm 0.1mHz, max}$.
However, for massive MBHs ($\gtrsim 10^7\msun$), the eccentricity distribution of
SBH-EMRIs is almost flat.

The eccentricity distributions of NS- and WD-EMRIs are similar to those of SBH-EMRIs, although they
are more concentrated towards the higher end.
Unlike for SBH-EMRIs, this is true even for very massive MBHs.
For all compact objects, the bias towards high eccentricities
is most pronounced around small MBHs (e.g., $\sim 10^5\msun$, see the bottom right panel
of Figure~\ref{fig:emris_eh}).

MSs can become EMRIs only if their masses are in the range of $0.1\msun\lesssim m_\star\lesssim1\msun$.
More massive MSs have larger physical sizes and are thus all tidally disrupted before becoming EMRIs.
Their eccentricity distribution is more concentrated towards low eccentricities compared to other stellar
objects.

BDs are more likely to become EMRIs if they are more massive.
For less massive BDs (e.g., $m\lesssim 0.05\msun$), their physical sizes are larger, and
their GW orbital decay is so weak that the critical distances $a_{\rm crit}$
defined in Section~\ref{subsec:critic_a} are much smaller.
Below the maximum eccentricity $e_{\rm 0.1mHz,max}$, their eccentricity
distribution peaks at lower eccentricities compared to compact objects.

The relatively high eccentricities of EMRIs observed in the LISA band can be understood as follows.
Most EMRI samples form in the regime
bounded by $T_{\rm GW}(x,j)\lesssim 0.1T_{\rm rlx}(x,j)$ and $j_{\rm lc}<j<1$.
Since stellar objects are more abundant in the outer parts of the cluster, within this bounded regime,
EMRIs are more likely to have higher eccentricities.
The details of the distribution are then regulated by the position of $a_{\rm crit}$,
the spin of the MBH, and the radial distribution of the stellar objects.

{Note that the eccentricities discussed here 
are defined within Newtonian physics (see more details in 
Appendix~\ref{apx:mapping}). A more accurate definition for eccentricity under relativistic 
framework should be 
given by the pericenter and apocenter of the relativistic orbit, e.g., $e=(r_a-r_p)/(r_a+r_p)$,
where $r_a$ and $r_p$ are the two roots of Equation~\ref{eq:Rr}. Usually the relativistic eccentricities are larger than those of Newtonian ones, as the pericenter can be closer than those in the Newtonian. 
Nevertheless, we still observe similar trends of bias towards high-eccentricity sources in the 
eccentricity distribution, which has been observed in some studies that adopt relativistic orbits 
and a Schwarzschild MBH~\citep[e.g.,][]{2005ApJ...629..362H, 2025arXiv250902394M}. 
We defer a full relativity study of the eccentricity and inclination distributions of EMRIs around 
non-spinning and spinning MBH to future works.
}
\subsubsection{The inclination distribution of EMRIs in the LISA band}
\label{subsec:evl_emri_inc}

\begin{figure*}
    \center
\includegraphics[scale=0.8]{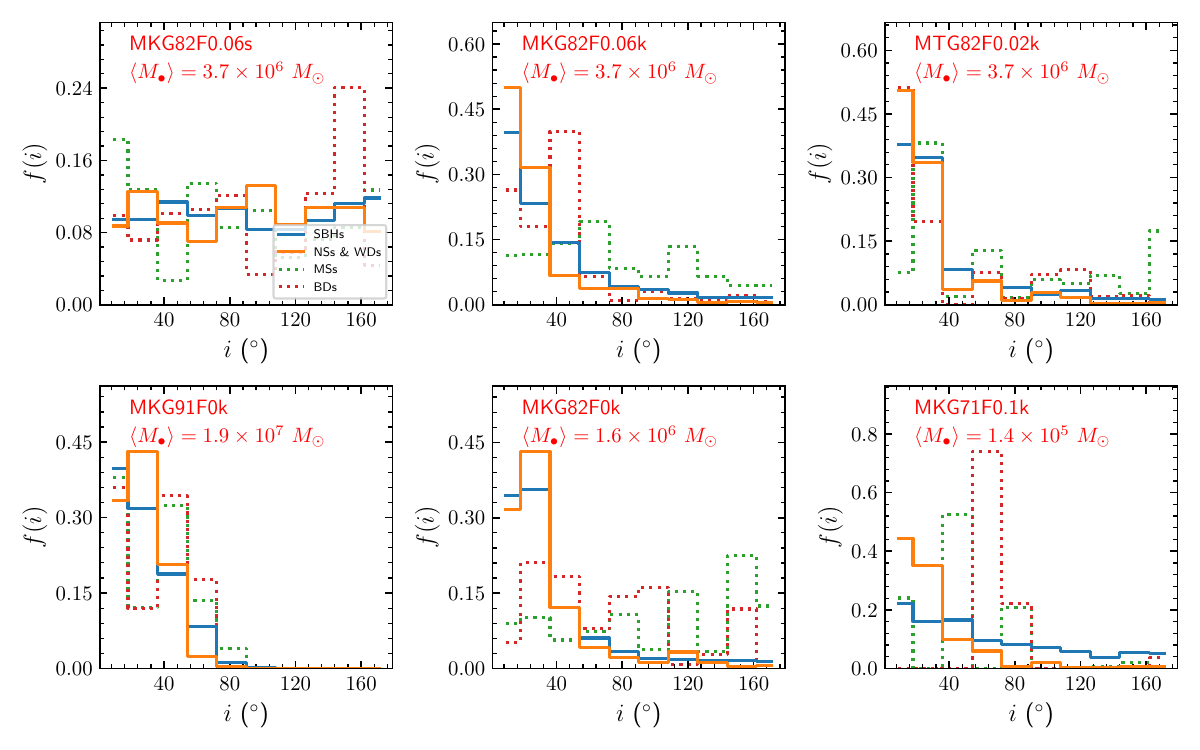}
\caption{Similar to Figure~\ref{fig:emris_eh} but for the orbital inclination distribution of EMRIs. }
\label{fig:emris_inc}
\end{figure*}

Figure~\ref{fig:emris_inc} shows the orbital inclination distribution of EMRIs.
We can see that if $a=0$, the inclinations of all EMRIs are uniformly distributed,
simply because in this case, the loss cone size is independent of the orbital inclination.

However, if $a=1$, EMRIs of compact objects are more likely to have
lower inclinations (i.e., they are more likely to be in prograde orbits).
The reasons are mainly twofold: (1)
ISOs with near-prograde orientations are smaller, so GW radiation is stronger and
stellar objects more easily become EMRIs; (2) If $\bh\gtrsim8\times10^6\msun$,
stellar objects in near-retrograde orbits will have already fallen into the MBH
before reaching $f_{\rm orb}=0.1$\,mHz.

For BD- and MS-EMRIs, it is possible that for a fraction of them, the loss cone is
determined by the ISO, while for the rest, it is determined by their tidal radii.
For the latter subset of samples, their inclination distribution should be
isotropic. Thus, for BD- and MS-EMRIs, the concentration of inclinations toward
lower angles is generally less pronounced. If $\bh\gtrsim 10^7\msun$, the loss cone of
most BDs is determined by the ISO (see the bottom right panel of Figure~\ref{fig:emris_me}),
and thus their inclination distribution is similar to that of compact objects.

\section{Discussion}
\label{sec:discuss}
By comparing the model results from Section~\ref{subsec:fix_component} and the results presented
in Section 4 of~\citetalias{Paper2} with those in Section~\ref{subsec:NSCs_sf},
we can see that ignoring stellar evolution leads to significant underestimates of the cluster's size expansion,
the loss of the NSC's total mass, and the mass growth of the MBH.
Thus, our results demonstrate that stellar evolution must be included to achieve realistic simulations
of NSC dynamics and MBH evolution, which is crucial for accurate predictions of TDE and EMRI properties.

Although our results provide substantial new insights into the evolution of NSCs, TDEs, and EMRIs,
several important limitations should be noted. First, our assumption of a single burst of star formation
in NSCs is an oversimplification, as observations suggest an
extended star formation history in both our Galactic Center~\citep[e.g.,][]{2010ApJ...708..834B} and
nearby galaxies~\citep{2020A&ARv..28....4N}.
However, as shown by ~\citet{2020NatAs...4..377N} and~\citet{2020A&A...641A.102S},
it is possible that $80\%$ of the current mass of the Milky Way's NSC formed between $11\sim 13$\,Gyr ago, suggesting
that our simplified scenario of single-burst formation $12$\,Gyr ago may not deviate substantially from reality.

Nevertheless, it will be necessary in the future to
incorporate star formation processes into our method, which will require
detailed modeling of the cosmic evolution
of the star formation history of galaxies, the conversion efficiency of gas to stars, and
the feedback mechanisms of the MBH's accretion activity.

Another limitation is our omission of the bulge and other mass components surrounding the NSCs.
For example, the current outer edge of a Milky-Way-like NSC in our simulation is around $\sim 1$\,kpc,
within which the enclosed dark matter mass can be up to $10^9\msun$, assuming a Milky Way NFW profile
\citep{2017MNRAS.465...76M}. Including these additional components may suppress
the further size expansion of the cluster, change the dynamics and leading to increase of
both the TDEs and EMRI event rates.

Additionally, we have ignored the natal kicks of NSs and SBHs during their formation.
\citet{2019MNRAS.484.3279P} suggest that natal kicks could eject a significant fraction of NSs from
an isolated NSC. However, many of them may remain bound to the system when considering
the additional gravitational potential created by the mass of the bulge and dark matter.
We defer these more complex scenarios to future studies.

Our study primarily focuses on the cosmological evolution of EMRI formation rates. This contrasts with analytical studies that often assume a steady-state distribution of stars to calculate the current population of EMRIs in the LISA band, such as the Early-EMRIs (E-EMRIs) studied by \cite{AmaroSeoaneMonoOligo} and the brown dwarf X-MRIs studied by \cite{2019PhRvD..99l3025A}. These analytical works highlight that the long residence time of these sources in the LISA band (from $10^5$ years for SBHs to $10^6$ years for BDs) can lead to a significant observable population even if the merger rates are low.

When comparing our time-dependent numerical results with these steady-state analytical predictions, we observe differences in the merger rates at 12 Gyr. The simulations show that long-term effects, such as stellar evolution mass loss and relaxation processes, lead to cluster expansion and a reduction in central density, which tends to suppress EMRI formation rates at late epochs compared to idealized analytical models.

However, the analytical models provide valuable estimates for the currently observable population, which is the relevant timeframe for the upcoming LISA mission. {The GNC simulations confirm the underlying dynamical mechanisms assumed in the analytical work. Provided some suiable conditions of the nuclear star cluster (such as the model M5\_2 in Figure~\ref{fig:emris_lc5}), the analytical estimates of the steady-state population (\cite{AmaroSeoaneMonoOligo}, \cite{2019PhRvD..99l3025A}) remain valid estimates (within an order of magnitude) for the present epoch}.
{These steady-state analysis offers a snapshot of the expected EMRI population.  as the astrophysical conditions are not expected to change significantly over the relatively short residence timescales of the EMRIs (e.g., the next $10^5$ years), even though the simulations demonstrate that these conditions evolve over a Hubble time.}

\section{Conclusion}
\label{sec:con}
Extreme mass-ratio inspirals (EMRIs) are important targets for next-generation space-borne
gravitational wave (GW) telescopes. Here, we investigate the dynamics and properties of EMRIs shaped by
the co-evolution of massive black holes (MBHs)
and nuclear star clusters (NSCs). We use our previously developed and well-tested Monte Carlo method,
\GNC, which can solve the self-consistent dynamics of NSCs with a mass-growing central MBH.

In order to investigate EMRI events, \GNC~has now been updated to include GW orbital decay,
the loss cone of a spinning MBH, and stellar evolution. We adopt
a Kroupa or top-heavy initial mass function (IMF) in the mass range of $0.01-150\msun$.
The mass of the MBH grows not only by directly swallowing objects that fall into the loss cone but also by
accreting gaseous material released by tidal disruption events (TDEs) and mass loss from stellar evolution.

We perform simulations over $12$\,Gyr, investigating
the density and size evolution of NSCs along with the mass growth of MBHs,
EMRIs of stellar-mass black holes (SBHs), neutron stars (NSs),
white dwarfs (WDs), main-sequence stars (MSs), and brown dwarfs (BDs),
as well as TDEs of MSs, BDs, and post-main-sequence stars (Post-MSs).

We observe a significant loss of the cluster's mass due to stellar evolution
($\sim40\%$ for a Kroupa IMF and $\sim75\%$ for a top-heavy IMF). The size
of the NSC expands due to mass loss and relaxation processes. However,
if the MBH is in a phase of Eddington-limited accretion and if the local
relaxation timescale exceeds the Salpeter timescale,
the size expansion can be slowed or even reversed.

Over $12$\,Gyr, the mass growth of the MBH contributed by TDEs
is typically $\sim 10^7\msun$, $\sim 10^6\msun$, and $\sim 5\times10^4\msun$
for massive (with a stellar mass of $10^9\msun$), Milky-Way-like, and
smaller NSCs (mass of $2\times 10^6\msun$), respectively.
The majority of MS-TDEs have masses of $\sim 0.5\msun$.
Meanwhile, the mass contributed from stellar mass loss depends critically on the assumed value of
$f_{\rm ma}$, which is the fraction of stellar mass loss that can be eventually accreted by the MBH.
Assuming $f_{\rm ma}\ge 0.1$, the final MBH mass is dominated by the contribution from
stellar mass loss. The MBH-to-NSC mass ratio can be up to $\sim 0.1$ ($\sim 0.8$)
if assuming $f_{\rm ma}=0.1$ ($f_{\rm ma}=1$).

The evolution of MS- and BD-TDEs exhibits similar decline-rise-and-fall behaviors discussed
in our previous work~\citep{Paper2}.
For Post-MSs, their TDE rates generally decline continuously over cosmic time.

Reproducing a Milky-Way-like NSC requires $f_{\rm ma}\lesssim 0.1$ and an initial effective size for the cluster
a few times smaller than the present value. This indicates that in our Galactic center, most
of the gas released by stellar mass loss is not accreted by the MBH.

We find that the evolution of EMRI event rates is mainly affected by the size expansion or contraction of the cluster,
the evolution of the stellar population, the IMF of the cluster, and
the spin of the MBH. Rapid mass growth of the MBH can slow or even reverse the
decline of EMRI rates. For BD- and MS-EMRIs, they exist only if the MBH mass
is much larger than $\sim 10^5\msun$; otherwise, they are more likely to be tidally disrupted.

Typically, the rates of SBH-, NS-, and WD-EMRIs peak at early epochs ($\lesssim 1$\,Gyr) at $10^{-7}-10^{-5}$\pyr,
$10^{-9}-10^{-7}$\pyr, and $10^{-8}-10^{-7}$\pyr, respectively,
and then gradually decline by a factor of $2-10$ over cosmic time.
As the populations of WDs increase continuously with time, their EMRI rates
decline more slowly than those of SBH- and NS-EMRIs.
MS- and BD-EMRI event rates typically increase with the mass of the MBH.
By $12$\,Gyr, their rates can be up to $10^{-8}\sim 10^{-7}$\pyr~ if the MBH mass can grow up to
$\sim 10^{7}\msun$.

For the models that can reproduce the Milky Way's NSC,
the EMRI rates of SBHs, NSs, WDs, BDs, and MSs are approximately
$10^{-7}$\pyr, $5\times10^{-9}$\pyr, $2\times10^{-8}$\pyr, $10^{-9}$\pyr,
and $4\times10^{-9}$\pyr at $12$\,Gyr, respectively.

We investigate the mass, eccentricity, and inclination distributions of EMRIs appearing in the LISA
band (when the orbital period reduces to $10^4$\,s). The masses of most SBH-, BD-, and MS-EMRIs are
in the ranges of $5\sim 23\msun$, $0.05\sim 0.1\msun$, and $0.1\sim 1\msun$, respectively.

For compact objects, they tend to have high eccentricities in the LISA band,
especially around Milky-Way-like or smaller MBHs ($\lesssim 4\times10^6\msun$).
If the MBH is maximally spinning, EMRIs are biased towards
low orbital inclination angles (prograde orbits).
In these two cases, the eccentricity and inclination distributions of both MS- and BD-EMRIs
are usually distinct from those of compact objects.

Our study provides numerous new details about the evolution of NSCs, the mass growth of MBHs, and EMRIs,
which will be useful for the analysis and understanding of EMRIs observed by LISA or other space-based GW observatories.
However, our conclusions are still limited by some oversimplifications in our simulations,
e.g., the omission of an extended star formation history in the cluster and
the bulge and dark matter components surrounding the NSC.
In the future, we will continue to incorporate more necessary recipes in
our study to achieve a more comprehensive and realistic understanding of NSCs over cosmic time.
\section{Acknowledgments}
\noindent
We thank Matteo Sadun Bordoni and Sebastiano Fellenberg for helpful discussions of this work.
This work was supported in part by the National Natural Science Foundation of
China under grant Nos. 12273006. This work was also supported in part by the Key Project of the 
National Natural Science Foundation of China under grant No. 12133004.
The simulations in this work were performed partly at the TianHe-II National
Supercomputer Center in Guangzhou.



	\appendix
	\renewcommand{\thefigure}{\thesection} 
\section{Numerical realization of gravitational wave orbital decay}
    \label{apx:GWnumerical}
\subsection{For orbits bound to the MBH}
    \label{apx:bound_orbit}

{For stellar orbits bound to the MBH, the decay of
energy and angular momentum per orbit}, $\overline{\delta E}^{\rm GW}$ and $\overline{\delta J}^{\rm GW}$,
are given by~\citep{Peters63,1964PhRv..136.1224P}:
\be\ba
&\overline{\delta E}^{\rm GW}=D^{\rm GW}_E P
=-\frac{64\pi}{5}\frac{\bh^2m(m+\bh)^{1/2}G^{7/2}}{r_p^{7/2}c^5}\frac{1+\frac{73}{24}e^2_K
+\frac{37}{96}e^4_K}{(1+e_K)^{7/2}},\\
&\overline{\delta J}^{\rm GW}=D^{\rm GW}_J P
=-\frac{64\pi}{5}\frac{\bh^2mG^{3}}{r_p^{2}c^5}\frac{1+\frac{7}{8}e^2_K}{(1+e_K)^2},
\label{eq:gw_emri}
\ea\ee
respectively, where $c$ is the speed of light, $m$ is the mass of the particle, and $P$ is the orbital period.

In reality, the GW dissipation of both $E$ and $J$ varies along the orbit,
with the maximum and minimum occurring near the pericenter and apocenter, respectively. For moderately eccentric orbits
(e.g., $e\lesssim0.9$), the variation is relatively mild, and thus the change in energy and angular momentum within a given time
$\delta t$ can be approximated by $D^{\rm GW}_E\delta t$ and $D^{\rm GW}_J\delta t$, respectively.

However, for highly eccentric orbits (e.g., $e\gtrsim0.9$),
GW dissipation is strongly concentrated near the pericenter, with negligible dissipation elsewhere along the orbit.
In these cases, applying a smooth damping term would lead to a significant overestimation of the dissipation
before the pericenter passage and an underestimation after it. This is particularly severe for particles weakly bound
to the MBH but with extremely high orbital eccentricity,
as before pericenter passage, the particle's dynamics may actually be dominated by relaxation rather than
GW radiation.

Therefore, for a particle with high eccentricity ($e>0.9$), in each orbital revolution,
we apply the full orbital changes $\delta E^{\rm GW}=D^{\rm GW}_E P$ and $\delta J^{\rm GW}=D^{\rm GW}_J P$ only if it has
passed the pericenter.

{
    This can be realized in our numerical simulations as follows:
    If $j>0.4$ ($e\lesssim 0.9$), we set 
    \be\ba
    \delta E^{\rm GW}=D^{\rm GW}_E \delta t, ~~&
    \delta J^{\rm GW}=D^{\rm GW}_J\delta t
    \ea\ee
    If $j<0.4$ ($e\gtrsim0.9$), we set
    \be\ba
        \delta E^{\rm GW}=D^{\rm GW}_EP n_p, ~~&
        \delta J^{\rm GW}=D^{\rm GW}_JP n_p
    \ea\ee
    where 
    \be\ba
    n_p&=\left\{\begin{array}{cc}
    k+1 & {\rm if~} \mathcal{M}_i<0~~{\rm and}~~\mathcal{M}_i+\delta \mathcal{M}\ge 0\\
    k & {\rm otherwise}\\
    \end{array}\right., ~~~
    \mathcal{M}_f=\mathcal{M}_i+2\pi k +\delta \mathcal{M}
    \ea\ee
    where $\mathcal{M}_i$ and $\mathcal{M}_f$ are the 
    initial and final mean anomaly of the orbit during a timestep $\delta t$, respectively.
}

{
    To ensure convergence we additionally 
place the following two constraints on the particle's
time step $\delta t$ (in addition to 
Equation B1-B3 of~\citet{Paper1}), such that 
\be\ba
\frac{\delta t}{P} &\le 
\left\{\begin{array}{cc}
    {\rm min} (\delta t_{\rm EGW},\delta t_{\rm JGW} ) & {\rm if}~j>0.4\\
    {\rm max} [{\rm min}(\delta t_{\rm EGW},\delta t_{\rm JGW} ), 0.2] & {\rm if}~j\le 0.4\\
\end{array}\right. 
\label{eq:deltatgw}
\ea\ee
where 
\be
\ba
\delta t_{\rm EGW}=\frac{0.01 E}{|D^{\rm GW}_E| },~~
\delta t_{\rm JGW}=\frac{0.005 J}{|D^{\rm GW}_J| }
\ea
\ee
When the eccentricity is high,
it is no longer necessary to decrease the time step as long as 
the pericenter passage is resolved. Therefore,  
we set a maximum of $0.2P$ on the 
time step in Equation~\ref{eq:deltatgw} when $j\le0.4$ ($e\gtrsim0.9$).
}

\subsection{For orbits unbound to the MBH}
    For stellar objects unbound to the MBH but bound to the cluster, after each
    pericenter passage, the decay of orbital energy and angular momentum are given
    by~\citep{1977ApJ...216..610T,1972PhRvD...5.1021H}:
    \be\bg
    \delta E^{\rm GW}=D^{GW}_E P=-\frac{64}{5}\frac{\bh^{2}m(m+\bh)^{1/2}G^{7/2}}{r_p^{7/2}c^5(1+e_K)^{7/2}}
    \times  \left[\left(\pi-\theta_0\right)\left(1+\frac{73}{24}e^2_K+\frac{37}{96}e^4_K\right)
    + \frac{(e^2_K-1)^{1/2}}{144}\left(301+\frac{673}{2}e^2_K\right)\right],\\
    \delta J^{\rm GW}=D^{GW}_J P=-\frac{64}{5}\frac{\bh^2mG^3}{r_p^{2}c^5(1+e_K)^2}
    \times \left[(\pi-\theta_0)\left(1+\frac{7}{8}e^2_K\right)
    +\frac{e_K}{8}\sin\theta_0(13+e^2_K)\right],
    \label{eq:gw_emri_hyper}
    \eg\ee
    respectively, where $\cos \theta_0=1/e_K$.
    Similarly to Appendix~\ref{apx:bound_orbit}, here we damp the energy and angular momentum
    only if the particle has passed the pericenter of its orbit.

    \section{Solving for the Relativistic ISO}
    \label{apx:relativistic_ls}

    \begin{figure*}
        \center
        \includegraphics[scale=0.6]{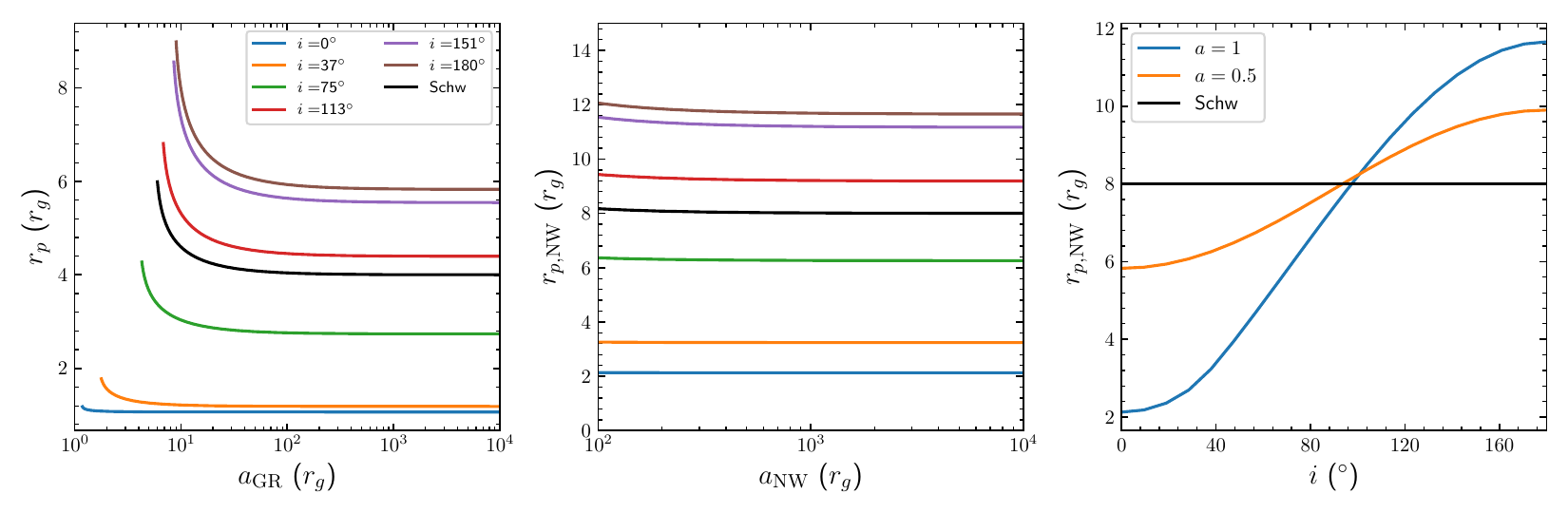}
        \caption{Left panel: The pericenter distance $r_{p}$ (in units of $r_g$) 
        of an innermost stable orbit (ISO) in Boyer-Lindquist (BL) coordinates, plotted as a function of the orbital semi-major axis, 
        defined as $a_{\rm GR}=(r_{p}+r_{a})/2$, where 
        $r_{a}$ is the apocenter distance. Both $r_p$ and $r_a$ are solved using the 
        method described in Appendix~\ref{apx:relativistic_ls}. 
        The colored lines show the results for the case of $a=1$ with various inclination angles $i$. 
        The solid black line shows the result for a Schwarzschild MBH. 
        Middle panel: The mapped Newtonian pericenter distance $r_{\rm p, NW}$ of the ISO 
        as a function of the Newtonian semi-major axis $a_{\rm NW}$, both defined by Equation~\ref{eq:aei}.  
        Right panel: The Newtonian pericenter distance $r_{\rm p, NW}$ of the ISO as a function of the Newtonian inclination angle $i_{\rm NW}$, 
        calculated for orbits with a semi-major axis of $a_{\rm NW}=10^4r_g$.
        }
        \label{fig:iso}
    \end{figure*}

    Consider a particle of mass $m$ orbiting a Kerr MBH of mass $\bh$.
    Its spacetime position is expressed in Boyer-Lindquist (BL) coordinates $(r,\theta,\phi,t)$.
    Let $r_a$ and $r_p$ denote the apocenter (maximum) and pericenter (minimum) radial distances to the MBH, 
    respectively. They are two of the roots of the following equation (in units where 
    $\bh=G=c=1$)~\citep{1972ApJ...178..347B}:
    \be
        R(r)=[(r^2+a^2)-a\lambda ]^2-(r^2-2 r+a^2)[\xi^2r^2+(\lambda-a)^2+q^2]=0
        \label{eq:Rr}
    \ee
    where $\xi=m/E$, $\lambda=L_z/E$, and $q^2=Q/E^2$. Here, $E$ is the energy, $L_z$ is the angular momentum component along the spin axis, 
    and $Q$ is the Carter constant of the particle. $a$ is the dimensionless spin parameter of the MBH. 
    For particles with $m\ne 0$, the above equation has four 
    real roots, $r_a\ge r_p\ge r_3>r_4$. For an ISO, the condition $r_p=r_3$ must be satisfied, which implies:
    \be
    \left.\frac{dR(r)}{dr}\right|_{r=r_p}=0.
    \ee
    The turning points in $\theta$ are given by the roots of the following equation:
    \be
    \Theta(\mu)=q^2-\mu^2\left[a^2(\xi^2-1)+\frac{\lambda^2}{1-\mu^2}\right]=0,
    \label{eq:Theta}
    \ee
    where $\mu=\cos\theta$. We define
    \be
    \cos i=\frac{L_z}{\sqrt{L_z^2+Q}}=\frac{\lambda}{\sqrt{\lambda^2+q^2}}
    \label{eq:spin_inc}
    \ee
    as the "inclination" of the orbit. One of the turning points in Equation~\ref{eq:Theta} then satisfies:
    \be
    \mu_+^2=\cos^2(\pi/2-i)
    \ee
    The value of $\lambda$ is given by the solution to the following equation~\citep{2011PhRvD..84l4060S}:
    \be
    [\mathscr{C},\mathscr{B}]\lambda^2+[\mathscr{D},\mathscr{B}]\lambda+[\mathscr{E},\mathscr{B}]=0
    \label{eq:lambda_eq}
    \ee
    where $[x,y]=x(r_a)y(r_p)-y(r_a)x(r_p)$ and 
    \be\ba
    \mathscr{B}&=-(r^2-2r+a^2)(r^2+\mu^2_+a^2)\\
    \mathscr{C}&=-\frac{a^2\mu_+^2+r^2-2r}{1-\mu_+^2}\\
    \mathscr{D}&=-4r a\\
    \mathscr{E}&=(r^2+a^2)(r^2+\mu_+^2 a^2)+2ra^2(1-\mu_+^2)
    \ea\ee
    $\xi$ can be solved for using:
    \be
    \mathscr{B}\xi^2+\mathscr{C}\lambda^2+\mathscr{D}\lambda+\mathscr{E}=0
    \label{eq:xi}
    \ee
    $q^2$ is then given by:
    \be
    q^2=\mu_+^2\left[a^2(\xi^2-1)\right]
    \label{eq:q2}
    \ee
    So far, for given values of $r_a$, $r_p$, and $i$, we can solve for $\xi$, $\lambda$, and $q^2$ 
    using Equations~\ref{eq:lambda_eq}, ~\ref{eq:xi}, and~\ref{eq:q2}.

    The remaining two roots of Equation~\ref{eq:Rr} are given by 
    \be
    r_3=\alpha+\sqrt{\alpha^2-\beta},~r_4=\alpha-\sqrt{\alpha^2-\beta}
    \label{eq:r34}
    \ee
    where 
    \be
    \alpha=-\frac{\xi^2}{1-\xi^2}-\frac{r_a+r_p}{2},~\beta=-\frac{a^2q^2}{1-\xi^2}\frac{1}{r_ar_p}
    \ee

    In general, the pericenter $r_p$ of an ISO for given values of eccentricity $e$, inclination $i$,
    and MBH spin $a$ can be solved numerically by the following steps:
    \begin{enumerate}
        \item Set the initial value of $r_p$ to $r_p=r_{\rm hz}=1+\sqrt{1-a^2}$.
        \item Calculate the apocenter distance $r_a=r_p/(1-e)$. Then, solve for $\lambda$ using Equation~\ref{eq:lambda_eq}.
        \item Solve for $\xi$ and $q^2$ using Equations~\ref{eq:xi} and~\ref{eq:q2}, respectively. Then, calculate $r_3$ and $r_4$ using Equation~\ref{eq:r34}.
        \item If $r_p$ is not equal to $r_3$, update $r_p$ using the rule $r_p\rightarrow {\rm max}[(r_p+r_3)/2, r_{\rm hz}]$ and return to step 2.
    \end{enumerate}
    The above steps are repeated until $r_p$ is sufficiently close to $r_3$. 
    The resulting value of $r_p$ is the pericenter of the ISO.

    For the case $a=0$ (a Schwarzschild black hole), $r_p$ and $\lambda$ can be expressed analytically as:
    \be
    r_p=\frac{2(3+e)}{1+e}r_g, ~\lambda={\rm sgn} (L_z) \frac{(3+e)^{3/2}}{2^{1/2}(1+e)^{1/2}}
    \ee
    Thus, in the limit $e\rightarrow1$, we have $r_p\rightarrow 4r_g$ and $|\lambda|\rightarrow 4$. For a circular orbit ($e=0$), $r_p=6r_g$.
    The left panel of Figure~\ref{fig:iso} shows the solution for $r_p$ for orbits around a Schwarzschild ($a=0$) or a maximally spinning ($a=1$) MBH.

    \section{Mapping to a Newtonian Loss Cone}
    \label{apx:mapping}
    The loss cone derived in the previous section is formulated in BL coordinates within a 
    relativistic framework. It cannot be directly applied to \GNC, 
    which is defined in a purely Newtonian context. 
    Consequently, a mapping between the relativistic and Newtonian loss cones must be established.

    {However, this mapping is not unique, as there is freedom in choosing 
    both the relativistic coordinate system and the mapping quantities}.
    For example, the transformation can be performed using the original BL coordinates, 
    the local non-rotating rest frame, or harmonic coordinates. {The 
    mapping quantities can also be defined by equating the conserved orbital quantities (e.g., energy and angular momentum), or by matching specific orbital elements (e.g., periapsis or apoapsis).}
    Nevertheless, all these mappings should converge at sufficiently large distances 
    from the MBH. As the majority of particles in \GNC~have apocenter distances $\gg r_g$, 

    {we perform the mapping between the Newtionan  and relativistic specific orbital energy 
    $E_{\rm NW}\rightarrow E/m-1=1/\xi-1$, $z$-component angular momentum $L_{NW,z}\rightarrow L_z/m=\lambda/\xi $ 
    and total angular momentum $L_{\rm NW}\rightarrow \sqrt{L_z^2+Q}/m=\sqrt{\lambda^2+q^2}/\xi$. The mappings 
    of orbital elements now become:
    \be
    \ba
    a_{\rm NW}&=\frac{1}{2}\frac{\xi}{\xi-1}\\
    e_{\rm NW}&=\sqrt{1-j^2}=\sqrt{1-(\lambda^2+q^2)\frac{2(\xi-1)}{\xi^3 }}\\
    \cos i_{\rm NW}&=\frac{L_z}{\sqrt{L_z^2+Q}}=\frac{\lambda}{\sqrt{\lambda^2+q^2}}=\cos i
    \label{eq:aei}
    \ea
    \ee
    }
{Given the values of $\xi$, $\lambda$ and $q$ for an ISO from the previous section},
    we can obtain the Newtonian pericenter $r_{\rm p, NW}=a_{\rm NW}(1-e_{\rm NW})$ 
    using Equation~\ref{eq:aei}.
    For nearly parabolic orbits where $e_{\rm NW}\simeq e\rightarrow 1$, the Newtonian angular momentum 
    approaches $L_{\rm NW}=(r_{\rm p,NW}(1+e_{\rm NW}))^{1/2}\rightarrow 
    (2r_{\rm p,NW})^{1/2}$, which in turn approaches $\sqrt{\lambda^2+q^2}$. 
    {For a Schwarzschild MBH, we can always choice coordinates such that $q=0$ and $L_{\rm NW}\rightarrow \lambda$. As} shown at the end of Section~\ref{apx:relativistic_ls}, $\lambda \rightarrow 4$ when $e\rightarrow 1$. 
    Thus, we immediately find that $r_{\rm p, NW}=8r_g$. 
    This is the commonly adopted value for the pericenter of an ISO around a Schwarzschild MBH in studies using Newtonian approximations.
    The solutions for $r_{\rm p, NW}$ for different MBH spin parameters $a$ 
    and orbital inclination angles are shown in Figure~\ref{fig:iso}.

    \end{document}